%% file: main.tex
\documentclass[sigconf, screen, authorversion]{acmart}
\AtBeginDocument{%
  \providecommand\BibTeX{{%
    \normalfont B\kern-0.5em{\scshape i\kern-0.25em b}\kern-0.8em\TeX}}}

\copyrightyear{2023}
\acmYear{2023}
\setcopyright{acmlicensed}\acmConference[UIST '23]{The 36th Annual ACM Symposium on User Interface Software and Technology}{October 29-November 1, 2023}{San Francisco, CA, USA}
\acmBooktitle{The 36th Annual ACM Symposium on User Interface Software and Technology (UIST '23), October 29-November 1, 2023, San Francisco, CA, USA}
\acmPrice{15.00}
\acmDOI{10.1145/3586183.3606813}
\acmISBN{979-8-4007-0132-0/23/10}
\acmConference[UIST '2023]{The ACM Symposium on User Interface Software and Technology}{Oct 29--Nov 1,
  2023}{California Bay Area, USA}
\acmSubmissionID{2679}

\input{meta/packages.tex}

\begin{document}

\input{meta/title.tex}
\input{meta/authors.tex}

\begin{abstract}
  \input{sections/00abs.tex}
\end{abstract}

\input{meta/keywords.tex}

\begin{teaserfigure}
  \includegraphics[width=\textwidth]{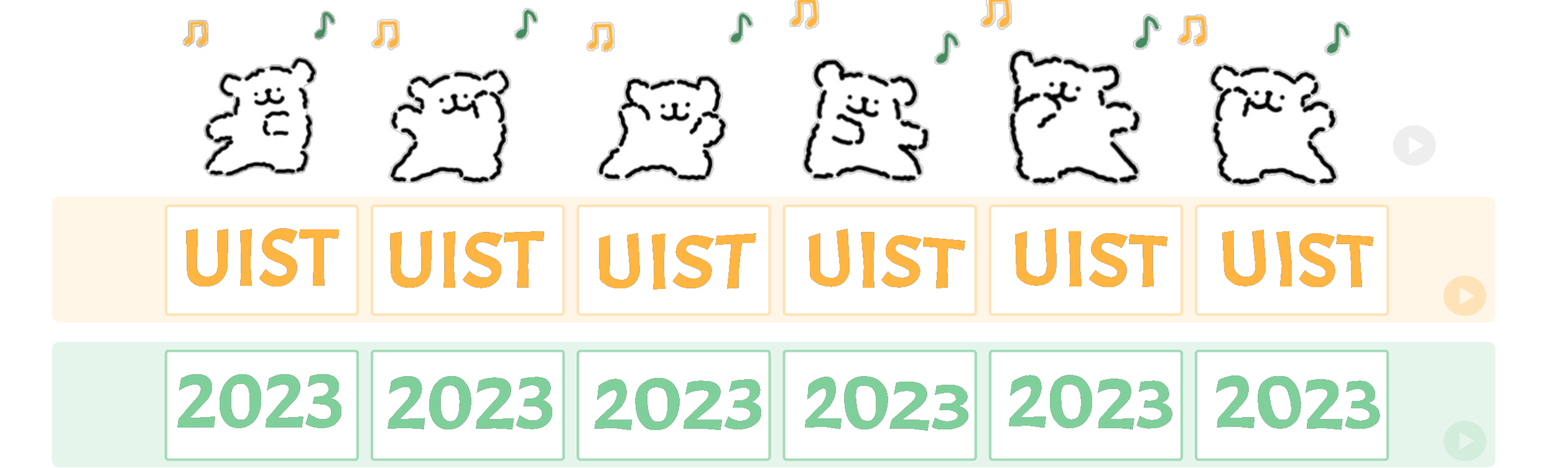}
  \caption{We introduce an approach to revive static text by transferring the motion of characters in a driving GIF.}
  \Description{Figure 1 is a row-wise illustration of animations. The first row is the driving GIF with a sketched puppy character arranged horizontally by frame. The second and third rows are also arranged by frame and depict corresponding movements in the texts “UIST” and “2023”, respectively.}
  \label{fig:teaser}
\end{teaserfigure}


\maketitle


\input{sections/01Intro.tex}
\input{sections/02RW.tex}

\input{sections/03Considerations}
\input{sections/04Methods.tex}

\input{sections/05Interface.tex}

\input{sections/06Implementation}

\input{sections/07Analysis}
\input{sections/08User}

\input{sections/09Discussion.tex}

\input{sections/10Conclusion.tex}

\input{sections/acks.tex}

\bibliographystyle{ACM-Reference-Format}
\bibliography{reference}

\input{sections/appendix.tex}

\end{document}

%% file: meta/packages.tex
\usepackage{xcolor}
\usepackage{booktabs}             
\usepackage{xspace}
\usepackage{listings}
\usepackage{wrapfig}
\usepackage{graphicx,calc}
\usepackage{enumitem}  
        
\newcommand{\eg}{\textit{e.g.}}
\newcommand{\ie}{\textit{i.e.}}
\newcommand{\ea}{{et~al.}\xspace}
\newcommand{\etc}{\textit{etc.}}
\def\tool{Wakey-Wakey}
\newcommand{\rev}[1]{{\color{black}{#1}}}

\setlist{noitemsep,topsep=0pt,parsep=0pt,partopsep=0pt}  
\newlength\myheight
\newlength\mydepth
\settototalheight\myheight{Xygp}
\AtBeginDocument{

}

%% file: meta/title.tex

\title{Wakey-Wakey: Animate Text by Mimicking Characters in a GIF}

%% file: meta/authors.tex

\author{Liwenhan Xie}
\authornote{Both authors contributed equally to this research.}
\email{liwenhan.xie@connect.ust.hk}
\orcid{0000-0002-2601-6313}
\authornote{This work is done during academic visit in Fudan University.}
\affiliation{%
   \institution{The Hong Kong University of Science and Technology}
 \city{Hong Kong SAR}
 \country{China}
}
\author{Zhaoyu Zhou}
\authornotemark[1]
\email{21210980099@m.fudan.edu.cn}
\orcid{0000-0002-8589-8088}
\affiliation{%
  \institution{Fudan University}
  \city{Shanghai}
  \country{China}
}

\author{Kerun Yu}
\email{19307110540@fudan.edu.cn}
\orcid{0009-0000-2647-4931}
\affiliation{%
  \institution{Fudan University}
  \city{Shanghai}
  \country{China}
}

\author{Yun Wang}
\email{wangyun@microsoft.com}
\authornote{Siming Chen and Yun Wang are the corresponding authors}
\orcid{0000-0003-0468-4043}
\affiliation{%
  \institution{Miscrosoft Research Asia}
  \city{Beijing}
  \country{China}
}

\author{Huamin Qu}
\orcid{0000-0002-3344-9694}
\email{huamin@ust.hk}
\affiliation{%
 \institution{The Hong Kong University of Science and Technology}
 \city{Hong Kong SAR}
 \country{China}}

\author{Siming Chen}
\authornotemark[3]
\orcid{0000-0002-2690-3588}
\email{simingchen@fudan.edu.cn}
\affiliation{%
\institution{Fudan University\\Shanghai Key Lab of Data Science}  
  \city{Shanghai}
  \country{China}
  }

\renewcommand{\shortauthors}{L. Xie, Z. Zhou, K. Yu, Y. Wang, H. Qu, and S. Chen}

%% file: sections/00abs.tex
With appealing visual effects, kinetic typography (animated text) has prevailed in movies, advertisements, and social media.
However, it remains challenging and time-consuming to craft its animation scheme.
We propose an automatic framework to transfer the animation scheme of a rigid body on a given meme GIF to text in vector format.
First, the trajectories of key points on the GIF anchor are extracted and mapped to the text's control points based on local affine transformation.
Then the temporal positions of the control points are optimized to maintain the text topology.
We also develop an authoring tool that allows intuitive human control in the generation process.
A questionnaire study provides evidence that the output results are aesthetically pleasing and well preserve the animation patterns in the original GIF, \rev{where participants were impressed by a similar emotional semantics of the original GIF}.
In addition, we evaluate the utility and effectiveness of our approach through a workshop with general users and designers.



%% file: meta/keywords.tex
\ccsdesc[500]{Human-centered computing~Graphical user interfaces}
\ccsdesc[500]{Computing methodologies~Animation}

\keywords{Kinetic typography, Animation, Motion transfer}

%% file: sections/01Intro.tex
\section{Introduction}
Nowadays, kinetic typography, \ie, animated text or motion text, has become common in daily life.
These vibrant artifacts can be observed in movies, website widgets, and online memes, \rev{such as the lyric video \textit{Skyfall}~\cite{skyfall} and the main title sequence of the movie \textit{Spider-Man}.}
Kinetic typography is effective for expressing emotional content, creating characters, and capturing or directing attention~\cite{shannon1998kinetic, lee2002engine}.
And there have been fruitful investigations of its application scenarios, including animated visualization~\cite{xie2023emordle}, instant messaging~\cite{gaylord2015body,kim2016yo}, ambient displays~\cite{minakuchi2008kinetic}, and captioning~\cite{Lee07EmotiveCaptioning}.

However, it remains non-trivial to craft the animation for text elements.
Leveraging commercial animation software~\cite{motion, aftereffects} or programming toolkits~\cite{lee2002engine} one may tweak the configuration of text in each animation keyframe,\eg, color, positions of the anchor point, and the transition between keyframes like a slow--in easing function.
Orchestrating these low-level parameters for a meaningful animation such as a melting scene requires careful considerations like which part of the text element to move, where to move, at what speed, \etc~  
As such, this process remains challenging and time-consuming with the large design space.
Previous studies~\cite{yeo2008kim, Lee07EmotiveCaptioning} tried to alleviate the authoring burden by designing a suite of templates.
While categorizing animated effects allows a one-click or even automatic generation, this approach suffers from limited customizability.
For instance, when one hopes to impress viewers with a refreshing presentation title, a pre-defined jumping effect may be inferior to a customized motion of breakdance.


Valuing uniqueness and personalization in digital communication~\cite{viegas09participatory, laura2013anytype}, we are motivated to find a sweet spot between automaticity and agency in kinetic typography tools.
Inspired by the recent advances in artificial intelligence, where a head image can talk by mimicking the motion of a driving video, \eg,~\cite{zhou2020makeittalk, hong2022depth}, we explore transferring existing animation designs to text.
The flourishing of GIFs on the web offers myriad high-quality animation references that imply emotions and humor, which can enrich the expressiveness of kinetic typography and make it easy for creators to specify desired effects.
However, existing approaches are not directly applicable to our goal.
On the one hand, research in motion transfer hardly attends to the non-photorealistic domains~\cite{siarohin2019first,xu2022motion}, especially for kinetic typography.
On the other hand, relevant research in text stylization focus on static text (\eg,~\cite{iluz2023word,xu2007calligraphic}), where the animation remains largely under-explored.



We propose a mixed-initiative framework for creating kinetic typography based on a driving GIF with a moving character.
On the machine side, the motion of the driving GIF is represented as the trajectories of several key points, which are extracted and guide the positional changes in the control points of the target text.
 On the human side, people can steer the mapping process by directly manipulating these points to refine the automatically computed positions of each point, resulting in a more desirable output.
 Based on the proposed framework, we develop an interactive interface for creating kinetic typography.
 We perform a series of evaluation studies to evaluate the usefulness and effectiveness of our approach.
 First, we demonstrate how individual components of the proposed framework contribute to the final result and test several cases.
 Second, a questionnaire study shows evidence that the output is both aesthetically pleasing and similar to the driving GIF.
 Third, we organize a workshop with general users and expert designers to evaluate the utility of our approach.

In summary, our work contributes to the following three aspects.
\begin{itemize}[leftmargin=2.2em]
    \item (Technique) An automatic approach to transfer the animation scheme from an anchor GIF to vector text.
    \item (Application) A prototype authoring tool for generating bespoke kinetic typography, which supports various scenarios, \eg, design prototyping and instant messaging.
    \item (Evaluation) A questionnaire study validating our transfer approach and a workshop demonstrating the usefulness of the authoring tool.
\end{itemize}

%% file: sections/02RW.tex
\section{Background \& Related Work}
In this section, we provide background information on typography and review existing research on kinetic typography, text stylization, and guided animation generation with an anchor.

\subsection{Preliminaries on Digital Typography}
Typography is defined as the art and technique of organizing text in a way that is easy to read, comprehend, and visually pleasing while presented.
In general, the visual appearance of a digital letter is determined by its \textit{font}, which is a particular size, weight, and style of a \textit{typeface}.
The typeface is a set of designed characters or letters, named glyphs, such as \rev{\texttt{Courier New}, {\fontfamily{ptm}\selectfont Times New Roman}, and {\fontfamily{pbk}\selectfont Bookman Old}}.
Internally, a typeface is represented in the raster domain or vector domain.
As bitmap fonts may become distorted or blurred with mosaic-like jagged edges at high resolution, we chose to adopt a vector-based typeface--the TrueType~\cite{penney1996truetype} font, which describes glyphs with quadratic bezier curves.

\subsection{Kinetic Typography}
Kinetic typography enriches animated user interface~\cite{chang1993animation} and digital media, which has received scholarly interest since the 1990s~\cite{shannon1998kinetic}. Most recently, Xie~\ea~\cite{xie2023emordle} summarized a design space of kinetic typography concerning changes in style, shape, position, and scale.
Compared with static text, kinetic typography is more competent in guiding attention~\cite{borzyskowski2004animated, minakuchi2008kinetic} and communicating emotions or semantics with the paralinguistic clues underlying animation~\cite{malik2009communicating, Lee07EmotiveCaptioning}.
Accordingly, there has been a series of works seeking to lower the burden of creating kinetic typography.
Kinetic Typography Engine~\cite{lee2002engine} set the basis of modern animation software (\eg, Adobe After Effects~\cite{aftereffects} and TypeMonkey~\cite{typemonkey}) with frame-based low-level specifications and a library of common effects.
The specification concerns text properties like position, rotation, \etc~
And the library was composed of functional time filters like oscillation.
TextAlive~\cite{kato2015textalive} featured kinetic typography synchronized with audio signals in video editing.

A stream of work investigated tools for average users rather than professional designers, where reducing efforts in animation configurations is a primary goal.
These works normally predefined a suite of animated effects and support selection or automatic matching under various contexts.
Instant messaging has been most studied,~\eg,~\cite{gaylord2015atim,yeo2008kim,forlizzi2003kinedit, minakuchi2005kinetic}.
For instance, Kinedit~\cite{forlizzi2003kinedit} allowed users to integrate text animation into a line of words.
Minakuchi and Tanaka~\cite{minakuchi2005kinetic} conceptualized an automatic composer that analyzes the semantics of text and queries suitable motions from a static repository to amplify its meanings.
Other scenarios include emotional animation for lyric videos~\cite{vy2008enact} and dynamic display based on viewers' emotions~\cite{lim2022study}.
These works suffered from \rev{the number of animated effects provided}.
For instance, there is hardly any consideration of transforming the text shape, which is common in animation~\cite{thomas1995illusion} yet requires by-frame editing.
In comparison, our work takes advantage of the ubiquitous online memes or stickers and can scrape their animation schemes to a random text with reliable transformation on its outlines.

\subsection{Text Stylization}
Our work closely relates to the task of text style transfer and semantic typography in text stylization, an area widely studied in computer vision/graphics to make a given text visually appealing.

Similar to our workflow, text style transfer concerns transferring the style of a given source (font samples, natural image/video) into text.
Some works explored propagating the design of a few stylized letters to others, such as typeface geometry~\cite{phan2015flexyfont} and glyph decorations~\cite{wang2019typography}.
Other works followed the general workflow of neural style transfer~\cite{gatys2016image} and viewed style as local neural patterns of the input image/video,~\eg, \cite{mao2022intelligent, yang2021shape, men2019dyntypo}.
In contrast, our work deforms the vectorized outline of the text to match the reference GIF.
We propose to vivify text by animating it in the way of a cartoon character, which diverges from their focus on learning image patch-based features.
Additionally, compared with kinetic typography, text style transfer emphasizes the artistic effect rather than an affective impact and typically produces static output.

Semantic typography amplifies the semantic meanings through visual cues in typography, which is also our goal.
Xu and Kaplan~\cite{xu2007calligraphic} proposed calligraphic packing, which deforms letters in a word to fit a given shape, which was improved by Zou~\ea~\cite{zou2016legible}.
In contrast to the intense deformation in letters, Word-As-Image~\cite{iluz2023word} stroke the balance of transformation on both sides, preserving the original font's style and legibility while ensuring the semantic implication, which was constrained by a pre-trained Stable Diffusion model~\cite{rombach2022high}.
Other approaches operate in the raster domain and leveraged external icons to replace parts of a text~\cite{tendulkar2019trick, zhang2017synthesizing}.
Our work differentiates from these works in that we imply semantics/emotion via animation of the text geometry rather than its static appearance, where the continuity between frames is considered.
To the best of our knowledge, this work is the first attempt to incorporate semantics in generating kinetic typography.


\subsection{Guided Animation Generation}
As we aim to produce emotionally or semantically resonant kinetic typography based on a given text, relevant constraints need to be introduced in the animation generation process.
Some works infer motions directly from a given still image, concerning features like texture~\cite{chuang2005animating, kazi2014draco, lai2016data}, status in a motion cycle~\cite{xu2008animating}, periodic patterns~\cite{halperin2021endless}, \etc~
These methods are unsuitable for our goal because a text usually appears with no background and is not equipped with equivalently rich properties for motion inference.

Motion transfer has been a standard task in computer vision, which is to generate a video based on a source image and a driven video by learning the motion from the driving video while preserving the appearance of the source image.
Monkey-Net~\cite{siarohin2019animating} was the first model-free approach to transfer motions of arbitrary objects by aligning key points between the source and target domain.
FOMM~\cite{siarohin2019first} further enhanced it with local affine transformations on the extracted key points.
It is one of the state-of-the-art models and we adapted it to fit the vector-based text.
Specifically, we maintained the text legibility by regularizing motion anchors with the distance change in the Laplacian coordinate.
Our method shares the same idea to preserve the structural information as DAM~\cite{tao2022structure}, which introduced a latent root anchor to model the structure of objects.
Different from our focus on the text, most existing datasets and models concern talking heads and human posture (\eg,~\cite{chan2019everybody, zhou2020makeittalk, siarohin2021motion, hong2022depth, smith2023tog}) and do not yield desired results on texts where legibility matters (see \autoref{sec:method}).
Our exploration of kinetic typography contributes to a unique case of cross-domain motion transfer.

In addition to fully automatic approaches, mixed-initiative interfaces for animation authoring have been investigated.
Users may specify the intended effect with sketch-based demonstration~\cite{kazi2014draco, xing2016energy, kazi2016motionamplifiers, willett2018mixed}, gestures~\cite{arora2019magicalhands}, or examples~\cite{dvoroznak2017example}.
Pose2Pose~\cite{willett2020pose2pose} supports creating cartoon character animation by minimizing the design efforts through clustering postures and automatically matching the stylized postures designed by the artists to the driving video. 
Most similar to our work, Live Sketch~\cite{su2018live} leveraged motion transfer to let novice users create animated sketches, where users are required to define control points in both the source and target domain. 
Our approach also allows users to specify their desired animation effect through a GIF, which is easy to access online.
However, the key points in the driving video are automatically extracted and automatically mapped to the target domain.
For a finer-grain control, users can  adjust the extracted key points and internal parameters.

%% file: sections/03Considerations.tex
\section{Design Considerations}
\rev{
Motivated to lower the barrier in creating kinetic typography, we explore motion transfer techniques. Instead of tweaking keyframe configurations from scratch, users may specify the desired animation effect based on a reference GIF. With numerous online GIF instances, users may derive more diverse animation effects compared with using template-based tools.

One major design consideration is to {\it support both direct generation and fine-grain refinement} (\textbf{C1}). We expect our approach can generalize to various user requirements, including casual use as in online-messaging and professional editing like video-making. In addition to producing one-off results, the tool should allow refinement over fine-grain configurations of each frame. This is because motion transfer inherently introduces uncertainties in the generated result from motion transfer, which may violate user preference.

Additionally, we strive to {\it empower creators with interpretable algorithmic parameters} (\textbf{C2}). We hope the system supports iterative refinement, which necessitates providing explanations for the generation process so that users can provide feedback and make adjustments at every step of the generation process. By combining human experience and supervision, we seek to achieve higher quality and more consistent generations in line with users' expectations.
}

%% file: sections/04Methods.tex
\begin{figure*}[h]
  \centering
  \includegraphics[width=\textwidth]{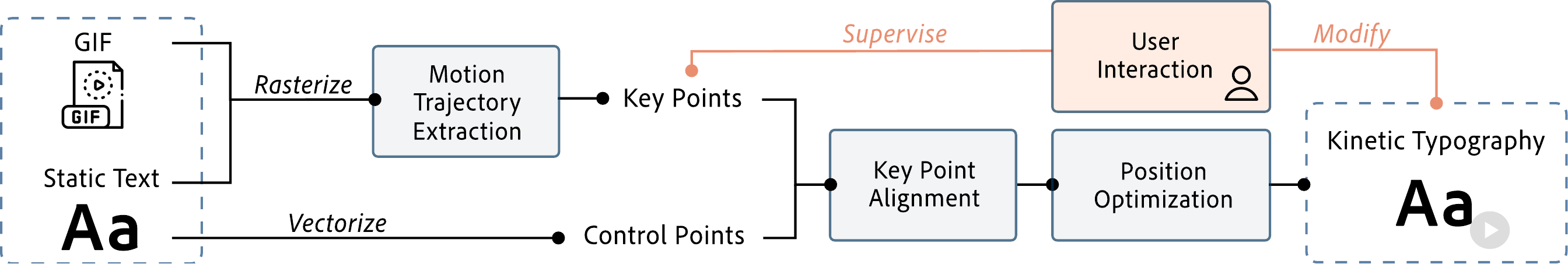}
  \caption{\rev{Overview of our approach. Inputs are a driving GIF and static text. The output is kinetic typography echoing the GIF's animations. The motion trajectory extraction module captures the key points in the driving GIF. The key point alignment module aligns the control points of the vectorized text to the key points. The position optimization module regularizes the text outline. And the User Interaction module allows human intervention on the intermediate key points and final results. }}
  \label{fig: overview}
  \Description{Figure 2 shows a conceptual diagram of the proposed method. The input is a GIF and a static text, and the output is kinetic typography. There are four modules. The motion trajectory extraction module receives the GIF and rasterizes static text as input and outputs the key points. This is input to the key point alignment module along with the control points obtained by vectorizing the static text, and output the deformed control points, which will be input to the position optimization module and output the final result. There is also a user interaction module pointing to key points and final control points for the adjustment.}
\end{figure*}
\section{Framework}
\label{sec:method}
In this section, we introduce a general framework for transferring the motion of a given GIF to a static text.
\subsection{Overview}
\autoref{fig: overview} illustrates our framework. 
\rev{The computational pipeline takes in a driving GIF and static text as input and outputs the kinetic typography. Users can tweak the intermediate key points and control points in the generated result (\textbf{C1}).}

Internally, the input text is represented in the TrueType format~\cite{penney1996truetype}.
It is first converted into an image and fed into a FOMM model~\cite{siarohin2019first} together with the anchor GIF to obtain the trajectory of the motion key points at each frame $X_i^f$, where the model identifies $N$ key points, and the GIF consists of $F$ frames, \ie, $i=1, ..., N$, $f=1, ..., F$.
The input text is also parsed to the initial control point set $C_j^0$ of its glyphs, with a total of $M$ control points, \ie, $j=1, ..., M$.
A local affine transformation is applied to both the initial control point set $C^0$ and the key point set trajectory $X^f$ to obtain the motion trajectory of the control point set $C_j^f$.
The updated control point trajectory $C_j^{\prime f}$ is attained through position optimization.
Subsequently, a vectorized glyph sequence is generated, culminating in the creation of animated text in vector form. 
Through the user interaction module, $X_i^f$ and $C_j^{\prime f}$ can be directly manipulated, and users can control some hyperparameters (\textbf{C2}).

\begin{figure*}[t!]
  \centering
  \includegraphics[width=\textwidth]{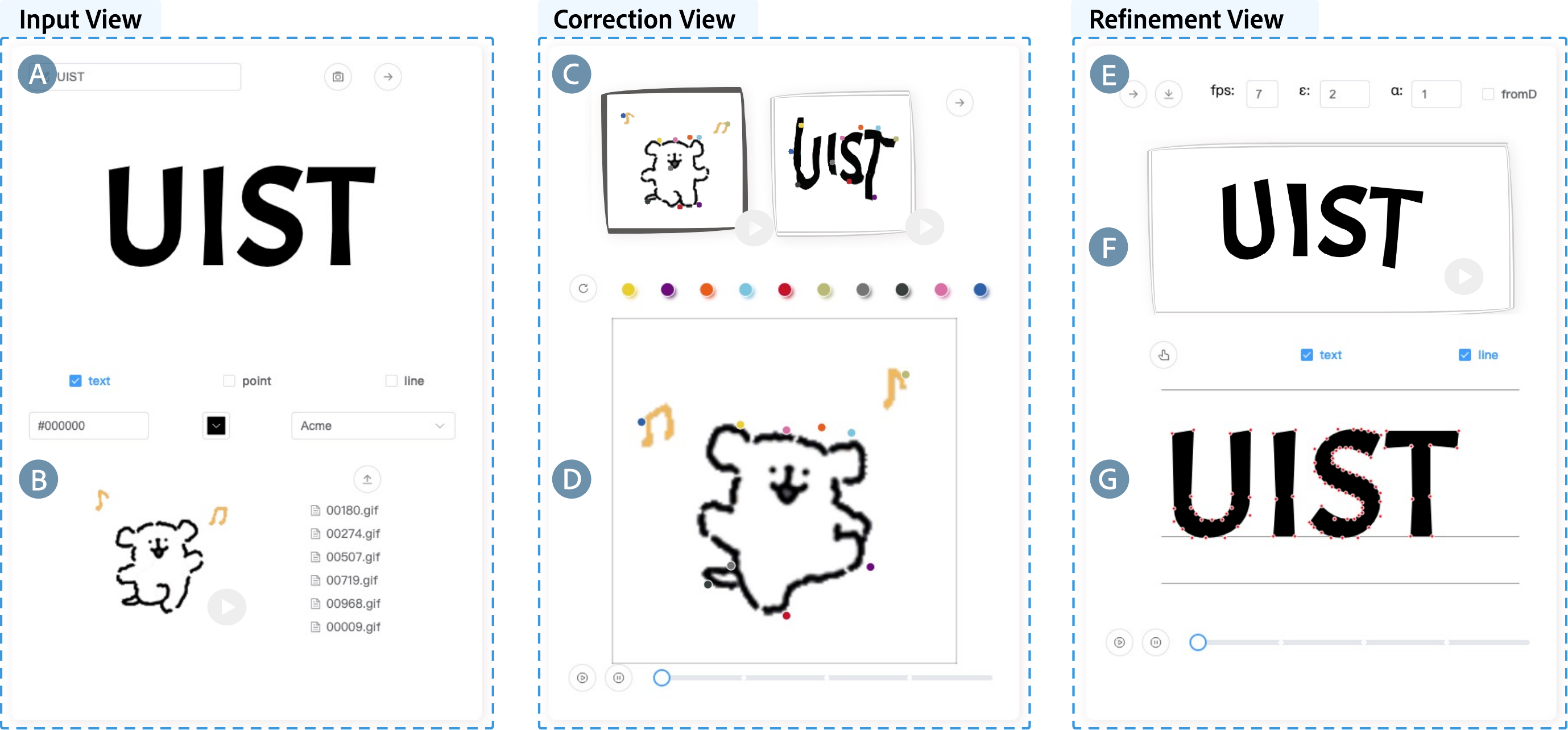}
  \caption{Wakey-Wakey: An authoring interface to interactively create anchor-based kinetic typography. There are three views: input view, correction view, and refinement view. (A) Input and preview the text, where font and color can be specified. (B) Upload a driving GIF. (C) Preview the matching of key points between the text and GIF at each frame. (D) Directly manipulate key points locations. (E) Fine-tune the hyperparameter. (F) Preview result GIF. (G) Refine the text control points at each frame. }
  \label{fig: interface}
  \Description{Figure 3 is a screenshot of the authoring interface for creating kinetic typography decomposed into seven areas. There are an input view, a correction view, and a refinement view. The input view allows for the input and customization of text (A) and the upload and preview of GIF (B). The correction view displays the key points in the animated image (C) and supports drag for correction (D). The refinement view contains parameter input boxes (E), supports control point editing (G), and displays the final result (F).}
\end{figure*}
\subsection{Motion Trajectory Extraction}
\label{subsec: fomm}
     We convert the static text to an image, and input it along with the anchor GIF to obtain the trajectory of motion key points. We adopted motion transfer to support the fast and flexible generation of kinetic typography.
As the object in the GIF usually differs from the text in shape, we need to separate the appearance and extract the motion trajectories of key points from the source GIF.

FOMM~\cite{siarohin2019first} is applied for key points extraction in our task. It is a self-supervised method using a framework that decouples appearance and motion, which effectively enriches the possible transferable motions to support motion transfer within any object category.
To address the problem of large differences in key points between the driving frame $\mathbf{D}$ and the source image $\mathbf{S}$, the FOMM model introduces an abstract reference frame $\mathbf{R}$ and obtains $\mathcal{T}_{\mathbf{S}\leftarrow \mathbf{D}}$ by separately calculating $\mathcal{T}_{\mathbf{S}\leftarrow \mathbf{R}}$ and $\mathcal{T}_{\mathbf{D}\leftarrow \mathbf{R}}^{-1}$.
$$
\mathcal{T}_{\mathbf{S} \leftarrow \mathbf{D}}=\mathcal{T}_{\mathbf{S} \leftarrow \mathbf{R}} \circ \mathcal{T}_{\mathbf{R} \leftarrow \mathbf{D}}=\mathcal{T}_{\mathbf{S} \leftarrow \mathbf{R}} \circ \mathcal{T}_{\mathbf{D} \leftarrow \mathbf{R}}^{-1},
$$
where $\mathcal{T}_{\mathbf{A}\leftarrow \mathbf{B}}$ denotes the mapping from the image $B$ to $A$.

In the implementation, $\mathcal{T}_{\mathbf{S}\leftarrow \mathbf{R}}$ and $\mathcal{T}_{\mathbf{D}\leftarrow \mathbf{R}}$ are obtained by key points detection in $\mathbf{S}$ and $\mathbf{D}$, respectively, which supports us to extract the key point trajectories from both the source and generated pixel-based text GIFs. Either of the two trajectories of the key points can be applied to drive the subsequent generation, and we use $X_i^f$ to represent the selected key point trajectory for simplicity. The separate detection mode also supports the relative generation ($\mathcal{T}_{\mathbf{S}_t\leftarrow \mathbf{S}_1}$ to deform from the source image) following a similar mindset, in addition to the absolute way ($\mathcal{T}_{\mathbf{S}_t\leftarrow \mathbf{D}_t}$ to deform from the corresponding frame of the source GIF directly).

In our implementation, we utilized the pre-trained FOMM model on the MGif dataset\rev{~\cite{siarohin2019animating}}, which has shown good performance in key point detection. 
\rev{Following the pre-trained model, the features extracted for each frame are estimated independently, and the number of key points is set to 10.}
However, to further enhance the integration of emotion into generation and analysis, we gather and create a dataset of Puppy Maltese~\cite{linedog} with 77 emotional-labeled GIFs and use it to fine-tune the model. This is done to better cater to the needs of the subsequent case studies and user surveys.

Due to the difficulty of FOMM in achieving good performance in motion transfer across different categories of objects, we only extract intermediate results from the key point detection module and redesign the subsequent generation steps based on our task.
We compare our result with the rasterized output of FOMM in a crowdsourcing study introduced in \autoref{sec:crowd}.

\subsection{Key Point Alignment}
\label{sec: keypointalignment}
A local affine transformation is applied to align the motion trajectories of key points to the control points. It introduces non-linearity to preserve local information better and achieve richer deformation. 
In our task, each key point extracted from each frame from the GIF is considered a local region, and the local affine transformation matrix set is obtained by computing the translational transformation of each key point in adjacent frames. The global nonlinear transformation is then calculated using a distance-weighted interpolation-based approach with the matrix set.
$$
\left[\begin{array}{c}
C_{j}^{f+1} \\
1 \\
\end{array}\right]
=\sum_{i=1}^{N} w_{i}(C_{j})\cdot\left[\begin{array}{c}
C_{j}^0 \\
1 \\
\end{array}\right] \left[\begin{array}{cc}
\mathcal{I} & 0 \\
(X_{i}^{f} - X_{i}^1)^{T} & 1
\end{array}\right],
$$
$$
w_i(C_{j})=\frac{1 / \|C_{j}^0 - X_{i}^1\|^{e}}{\sum_i 1 /\|C_{j}^0 - X_{i}^1\|^{e}}.
$$
$C_{j}^f$ and $X_{i}^f$ denote the control point $j$ and the key point $i$ at frame $f$, respectively. The control point's position at each frame is calculated in reference to the key point at the first frame to achieve global stability. $\mathcal{I}$ is a 2nd-order identity matrix. $w_i$ is a weight function for a control point with respect to the key point $i$.

The weight decays according to the inverse of the $e$-th power of the relative distance from $X_i$ to $C_j$, where $e$ controls the locality of the affine transformations, \ie, the degree to which each affine transformation affects the target point.
\begin{figure*}[t]
  \centering
  \includegraphics[width=\textwidth]{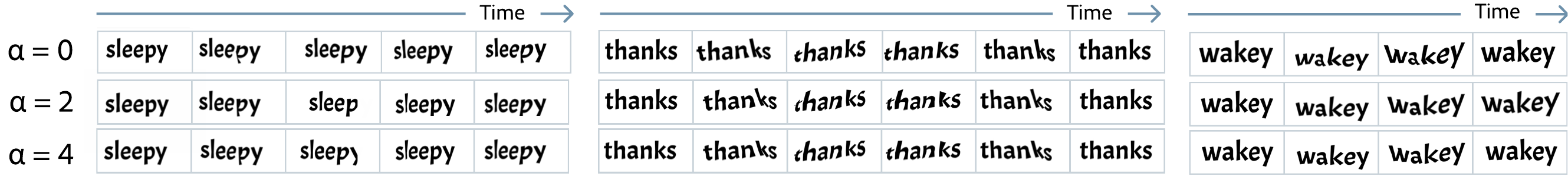}
  \caption{Comparison of the generation results with $\alpha = 0, 2, 4$. Increasing $\alpha$ enhances the smoothness of the glyph, but an excessive value of $\alpha$ may negatively impact the amplitude of the motion.}
  \label{fig: ablation1}
  \Description{Figure 4 contains three blocks showing cases applying different parameter alpha. The vertical axis represents different values of the parameter alpha (0, 2, 4), and the horizontal axis shows the generated kinetic typography from three words "sleepy", "thanks", and "wakey" arranged by frame.}
\end{figure*}
\subsection{Position Optimization}
To alleviate inappropriate deformation of glyphs caused by changes in the relative position of the control points, we optimize the positions of the control points by frame based on the Laplacian coordinate, which generally describes the relative positions on the surface using the neighbor information.
For a control point $j$ at frame $f$, its Laplacian coordinate $L_j^f$ is calculated as
$$
\begin{aligned}
    L_{j}^f &= \sum_{k \in N_{j}}\omega_{jk}^f\left(C_{k}^f-C_{j}^f\right) =\sum_{k \in N_{j}}\omega_{jk}^f C_{k}^f - C_{j}^f,
\end{aligned}
$$
where $C_j^f$ and $C_k^f$ denote the Cartesian coordinate of the control point $j$ and $k$, respectively. $N_{j}$ is the set of $K$-nearest neighboring control points with the smallest Euclidean distance to the control point $j$, which is calculated based on the initial control points set $C^0$, invariant to changes in $f$. $\omega_{jk}^f$ denotes the weight of the neighbor point $k$ in the Laplacian representation of the control point $j$, where
$$
\omega_{jk}^f=\frac{1 / \|C^{f}_{k} - C^{f}_{j}\|^{2}}{\sum_{k \in N_j} 1 /\|C_{k}^f - C_{j}^f\|^{2}}.
$$

Considering the inhomogeneity of the distribution of discrete sampling points, we use the aforementioned distance-based weights to describe the detailed location information better.
Further, we design the following objective function  $\mathcal{L}_{\text{total}}$ to optimize the coordinates of the sequence of control points obtained by frame.
$$
\begin{aligned}
 \mathcal{L}_{\text{total}} &= \alpha \cdot  \mathcal{L}_{\text{glyph}} + \mathcal{L}_{\text{motion}}, \ \alpha\in[0,+\infty),\\
 \mathcal{L}_{\text{glyph}} &= \sum_{j=1}^{M}\left\|L_{j}^f-L_{j}^0\right\|^{e},\  
 \mathcal{L}_{\text{motion}} = \sum_{j=1}^{M}\left\|C_j^f - C_j^{f\prime}\right\|^{e}.\\
\end{aligned}
$$

$L_{j}^f$ and $L_{j}^0$ denote the Laplacian coordinates of the control point $j$ in frame $f$ and $0$, respectively. $C_j^f$ and $C_j^f\prime$ denote the coordinates of the control points before and after optimization.
$\mathcal{L}_{\text{glyph}}$ measures how much the local shape details are preserved, which is computed as the sum of the distance of Laplacian coordinates between the optimized and initial control points. 
$\mathcal{L}_{\text{motion}}$ measures how much of the motions are preserved, as the sum of the distance of the control points before and after optimization, i.e., minimizing edit distance.
$\alpha$ is a hyperparameter representing the trade-off between the two loss functions.
The larger $\alpha$ is, the more details of the initial glyph and the less motion are preserved.
As for the norm $e$, the larger it is, the locality is more regulated, which leads to stronger deformation.
We use a $K$-dimensional tree to accelerate the nearest-neighbor search, where the parameter is empirically set: $K=3$.
\rev{While we employ frame-by-frame optimization, we note it is worth introducing global temporal regularization terms in the loss function to promote smoothness and consistency. 
}

\subsection{User Interaction}
The user interaction module allows direct manipulation of the computed positions of the key points and control points at each frame, \ie, $\{X_i^f\}$ and $\{C_j^f\}$, $\forall i \in [1, N]\cap \mathbb{N}$, $j\in [1,M]\cap\mathbb{N}$, $f\in [1, F]\cap\mathbb{N}$.
In this way, creators of kinetic typography can participate in the motion transfer process and adjust the final results according to their needs.
The hyperparameter $\alpha$ in the position optimization stage can also be adjusted for different texts, as illustrated in \autoref{sec:ablation}.

%% file: sections/05Interface.tex
\section{Authoring Interface}
\label{sec:authoring_tool}
Based on the proposed framework, we implement a mixed-initiative authoring tool called Wakey-Wakey\footnote{\rev{The name suggests that the authoring tool awakes static text and makes it lively.}} that allows fine-grain adjustment for more natural and aesthetic results (see \autoref{fig: interface}).
This section offers a step-by-step guide showing how to interactively generate animated text with our tool referring to the interface.


The user-oriented process mainly consists of three steps: text and GIF input, key point correction, and glyph refinement, each supported by a view: \textit{Input View}, \textit{Correction View}, and \textit{Refinement View}.
While the input step is mandatory, the correction and refinement stages are optional. This allows for simple end-to-end personalized generation, as well as interactive improvement.

\paragraph{Input View.}
Users first input the text and customize its static appearance with the global typeface and color (\autoref{fig: interface} A).
They can upload and preview the driving GIF through a button (\autoref{fig: interface} B) .
The section will record and list the recent upload history. 
After clicking ``Next'', the system will process the input with our method, and both intermediate and final results will be displayed in the other two views. Users can easily obtain the generated animated text here without any additional effort.

\paragraph{Correction View.}
Users can then drag the displayed the key points from the motion trajectory extraction module to a suitable location at a specific frame, as shown in \autoref{fig: interface} D.
Special attention can be paid to the key point trajectory with a corresponding colored button above. 
Two thumbnails (\autoref{fig: interface} C) enable switching between the anchor GIF and the extracted key points for correction.
As the corresponding key points share the same color, users can learn how the mapping is.
Users are supported to ensure better quality by maintaining reasonable motion trajectory to drive the generation of the vector animated text.

\paragraph{Refinement View.}
Users may configure the parameters $\alpha$ and $e$, and select whether the kinetic typography is generated by aligning to key points from the anchor GIF or the extracted key points (\autoref{fig: interface} E).
The bottom panel (\autoref{fig: interface} G) enables the users to adjust the glyph by dragging the control points.
And the final GIF of kinetic typography is displayed in the middle panel (\autoref{fig: interface} F).

%% file: sections/06Implementation.tex
\section{Implementation}
\rev{\tool{}\footnote{Source code available at \url{https://github.com/KeriYuu/Wakey-Wakey}.} was implemented as a client/server web application.
The front end was built with \texttt{Vue} for user interactions.
The computational framework for generating kinetic typography was implemented in Python.
An automatic generation takes around 300ms/frame (CPU: Intel i7 4.9 GHz).
The \texttt{Flask} framework is used to handle the messaging between the front end and the back end.
}

%% file: sections/07Analysis.tex
\section{Method Analysis}
\label{sec:crowd}
Due to the absence of pre-defined ``ground truth'' and lack of standard metrics in the nascent area of motion transfer for quantitative assessment, we empirically evaluated our approach by (1) analyzing the impacts introduced by each component, (2) comparing the automatically generated result from different styles of GIFs and typefaces, and (3) conducting questionnaire studies to understand how general people perceive the outputs based on several cases.

\subsection{\rev{Effects of Components}}
\label{sec:ablation}
\rev{We evaluated the effect of each component in the workflow} to analyze how our adaption to FOMM and the introduced human interventions can improve the generated result, including local position optimization, vectorized text representation, key point correction, and glyph refinement.

\paragraph{Local Position Optimization}
The position optimization module is introduced to preserve the local shape of each glyph better.
\autoref{fig: ablation1} demonstrates three motion transfer results with $\alpha$ set to $0, 2, 4$. 
$\alpha$ is the weight of $\mathcal{L}_{\text{glyph}}$, which controls the degree of preservation of local shape. As can be seen, when alpha is set to 0, i.e., without local position optimizing, some local parts of the glyphs are unsatisfactory, such as the ``p'' in ``sleep'', the ``t'' and ``k'' in ``thanks'', and the ``a'' and ``k'' in ``wakey''.
With the increment of $\alpha$, the glyph becomes smoother. However, when alpha is too large, it may cause too much preservation of the original glyph and result in a loss of motion, for example, the ``w'' and ``k'' of ``wakey'' when $\alpha = 4$. Through experiments, we find a suitable default value of 2.

\label{sec:case}

\begin{figure}[t]
  \centering
  \includegraphics[width=\linewidth]{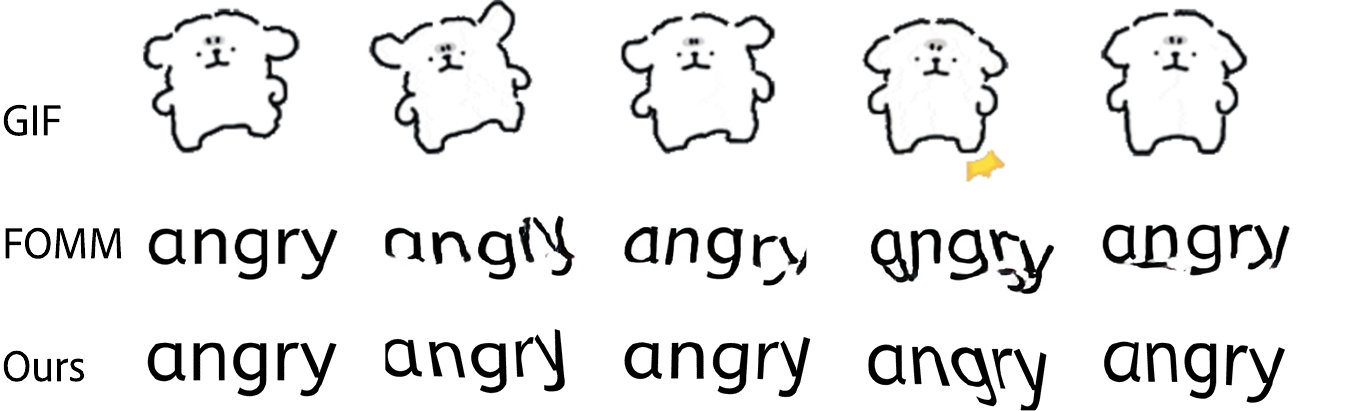}
  \caption{A comparison with the FOMM model~\cite{siarohin2019first}. Our approach operates on the control points of vectorized text, which improves legibility.}
  \label{fig: case}
  \Description{Figure 5 contains three rows of animation keyframes. From top to bottom, they are the driving GIF, the rasterized animated text generated by the FOMM model, and the vectorized kinetic typography generated by our method.}
\end{figure}
\paragraph{Vectorized Text Representation.}
Instead of directly employing existing image-based motion transfer models, our approach operates on the control points of text glyphs.
As shown in \autoref{fig: case}, the output results based on pixels are not stable enough.
For example, in the pixel-based glyph generated by FOMM, the letter ``y'' of ``angry'' has breaks and extra noisy strokes.
As the anchor GIF is hardly the targeted category of kinetic typography, using FOMM for motion transfer does not produce satisfactory results. 
In contrast, our method better preserves the integrity and legibility of the glyph and produces a more stable frame sequence.

\begin{figure}[h]
  \centering
  \includegraphics[width=\linewidth]{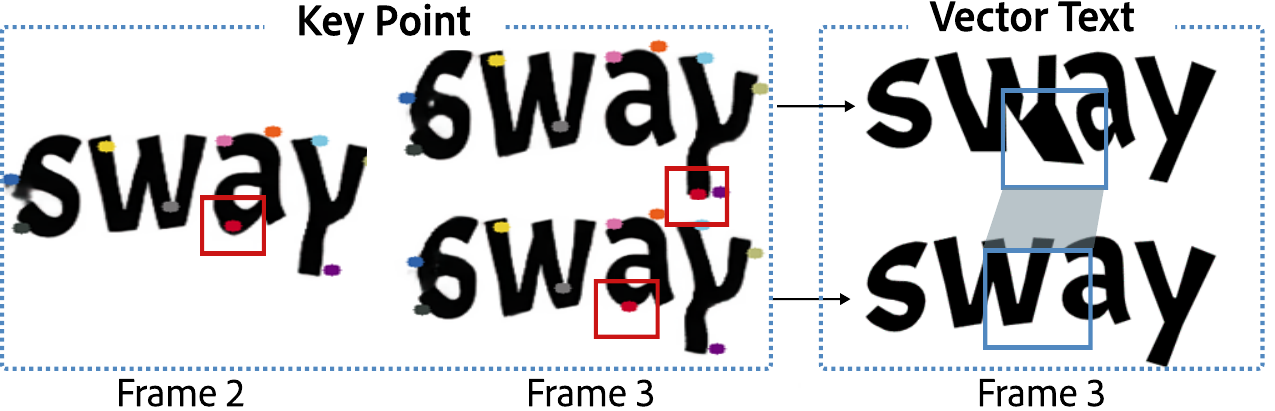}
  \caption{Comparison of the output before (top) and after (bottom) key point correction. The automatic mapping fails when the highlighted red key point shifts to another location in two consecutive frames, which yields distortion in the text.}
  \label{fig: ablation2}
  \Description{In Figure 5, the left part shows the key points before and after correction, and the right part shows the corresponding generated vector text.}
\end{figure}

\paragraph{Key Point Correction.}
The key points $\{X_i^f\}$ detected by the model may not always be accurate, which can result in unexpected deformations in the generated animated text that rely on these key points. 
Our approach allows users to interactively correct the key points, thereby obtaining a more desirable generation that aligns with their expectations.
As shown in \autoref{fig: ablation2}, by analyzing the preceding and following frames, we can find that in the fourth frame of the pixel-based animated text image sequence generated by FOMM, the key point marked in red noticeably shifts towards the right. 
This caused an excessive deformation towards the right in the lower right part of the letter ``W'' in the vector-based animated text generated with this key point. 
By dragging the key point towards the left to an area consistent with the preceding and following frames, the deformation of the generated glyph appears more reasonable and smoother.

\begin{figure}[h]
  \centering
  \includegraphics[width=\linewidth]{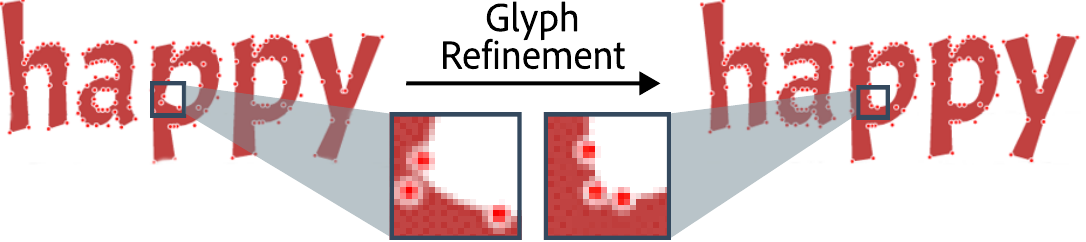}
  \caption{Manual refinement of the text control points. By dragging the distorted control points, one may intuitively refine the output kinetic typography at a fine-grained level.}
  \label{fig: ablation3}
  \Description{In Figure 7, the left and right are the glyphs before and after the glyph refinement, and the center shows the control points that were adjusted.}
\end{figure}

\paragraph{Glyph Refinement.}
The control point sequence $\{C_j^f\}$ can be manually updated for fine-grain refinement.
Through the authoring interface, users are supported to drag the control points and preview the result immediately.
As shown in \autoref{fig: ablation3}, the sharp corners inside the first letter ``p'' affect the glyph aesthetics, where the highlighted left line segment is tilted to the left and needs to be adjusted. 
By moving the three control points in the sharp corner area to the right and adjusting the relative positions of the three points, the refinement process is done. 
It is evident that the updated glyph achieves a better effect through simple and immediate dragging.

\subsection{\rev{Generalizability}}
\label{sec:generalizability}
\begin{figure*}[h]
  \centering
  \includegraphics[width=0.8\textwidth]{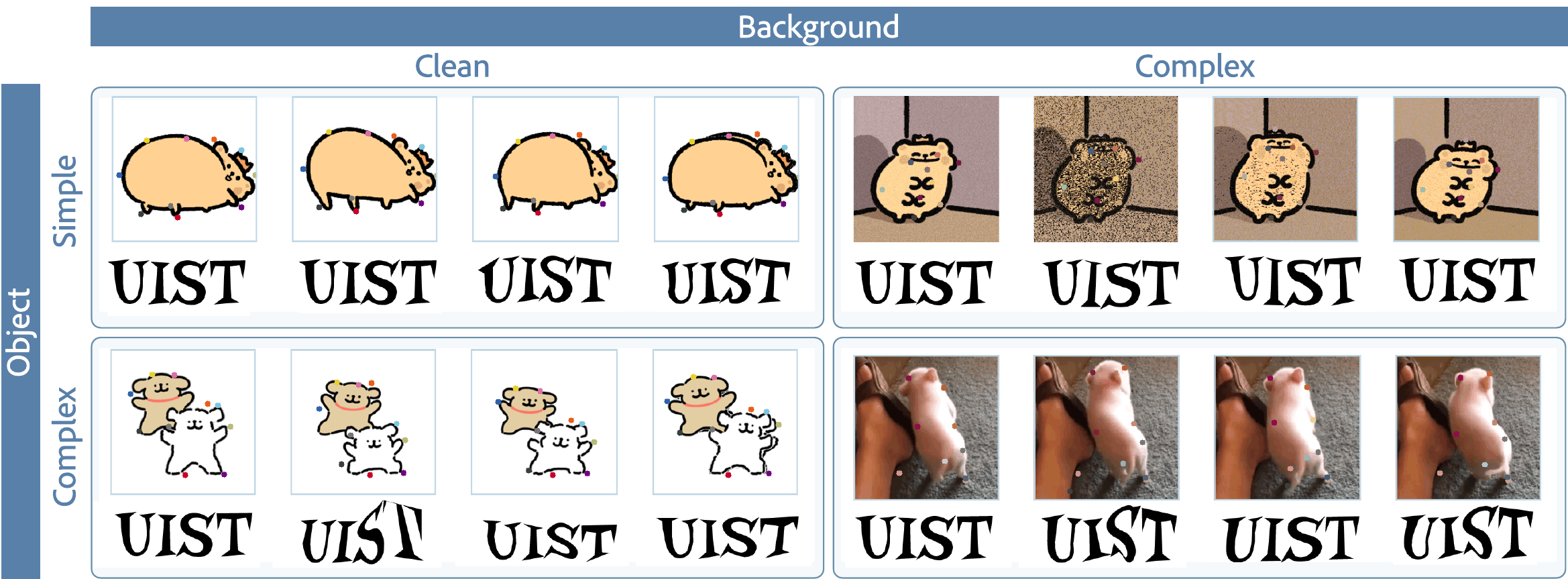}
  \caption{\rev{A comparison of results from driving GIFs of different complexities in background and target object(s).}}
  \label{fig: complexity}
  \Description{In Figure 8, there are four blocks displaying keyframes of the generated results of “UIST” from different complexities of GIF inputs.}
\end{figure*}

\begin{figure*}[h]
  \centering
  \includegraphics[width=\textwidth]{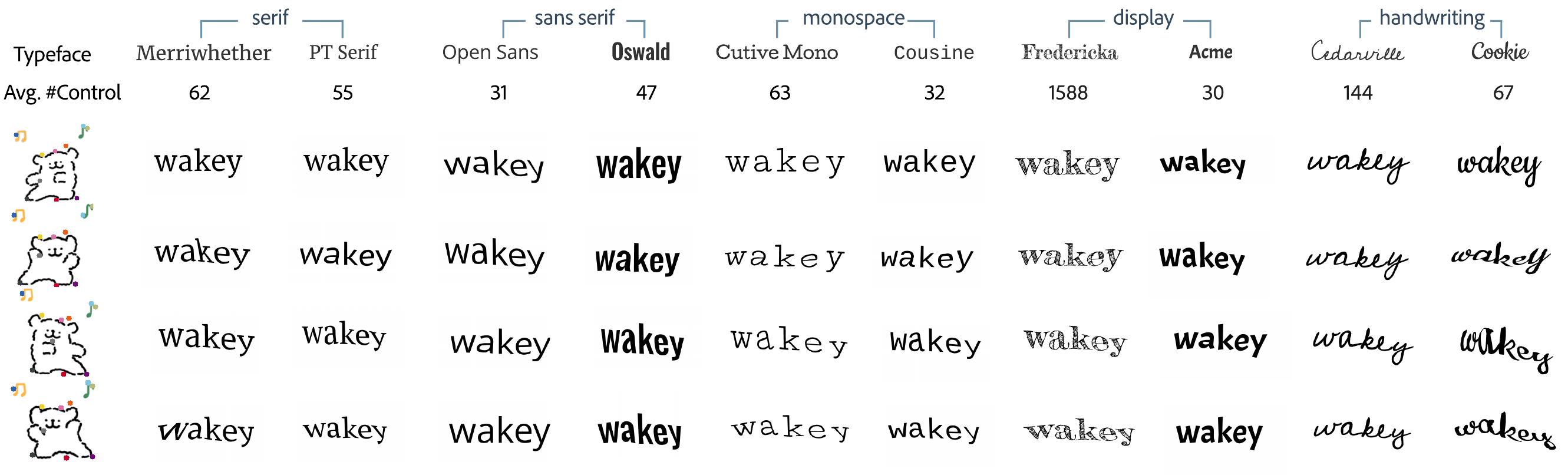}
  \caption{\rev{A comparison of results from typefaces of different categories and average number of control points for the 26 English alphabets (Avg. \#Control). The first column shows four key frames of the driving GIF. Each column in the rest shows the font information and the corresponding motion transfer result.} }
  \label{fig: font}
  \Description{In Figure 9, there are eleven columns. The first column shows four keyframes of the driving GIF. The rest ten columns show corresponding keyframes based on different types of typefaces. Three additional rows in each column present the class of the typeface, the name of the typeface, and the average number of control points for 26 English alphabets in the typeface.}
\end{figure*}

\rev{
Drawing from our experience, we reflect on the generalizability of our approach in terms of the driving GIF and the input typeface.

In general, \tool{} can accommodate input GIFs with a clean background and a simple-shape moving rigid body, such as instances from the Puppy Maltese dataset.
This is because our implementation adopts the pre-trained FOMM model based on the MGif dataset, which features a white background and one cartoon animal. 
Seen from \autoref{fig: complexity}, the automatic key point extraction may fail and cause large distortion when there are multiple moving objects, or the moving object exhibits complex patterns.
A complex background also threatens the reliability of extracted key points in the driving GIFs, such as a clip from natural videos.
While these issues can be addressed by manual correction, we also note that the motion trajectory extraction module can be improved by unsupervised training on a larger dataset with representative cases or using a large universal model.

As for the input fonts, our approach empirically performs well for typefaces with more than 5 control points in a glyph.
The more control points encapsulated in the typeface, the more likely that the position optimization can maintain its legibility.
\autoref{fig: font} showcases the automatic generation results for ten typefaces of common classes.
Typefaces with the most control point number also yield the most smooth results, including Fredericka and Cedarville.
However, there might be strong deformation for handwriting-styled typefaces, potentially due to their high flexibility.

}

\subsection{Questionnaire Study}

We conducted two questionnaire studies to evaluate the effectiveness of our approach.
Specifically, we seek to understand (1) whether our approach convincingly transfers the motion, and (2) to what extent the semantics of the original GIF can be preserved.

\subsubsection{Setup} The questionnaires are distributed on Qualtrics.
\rev{Participants are required to complete Study I before Study II}. And the questions appear in a random order in each study. 
We used meaningless pseudo-words from the Lorem Ipsum corpus~\cite{lorem} as input text in order to minimize the influence of text content.
\rev{For driving GIFs, we used the Puppy Maltese dataset to generate cases. To avoid confounding effects, the driving GIFs are non-repetitive. And we employed the typeface ``Akronim'' for it has over 300 control points, which may lead to satisfying results without human intervention and therefore suitable for our scenario requiring batch generation.}

\paragraph{Study I: Motion Transfer}
The first study aimed to evaluate the overall quality of the output kinetic typography.
As no quantitative metric is available in our task, we obtained subjective assessments by asking the participants to rate the similarity between the driving GIF and the output kinetic typography and their aesthetics.
On the one hand, the similarity between the source and target is the primary goal in motion transfer.
On the other hand, aesthetics is a common pursuit in animation design.

\begin{figure*}[h!]
  \centering
  \includegraphics[width=\textwidth]{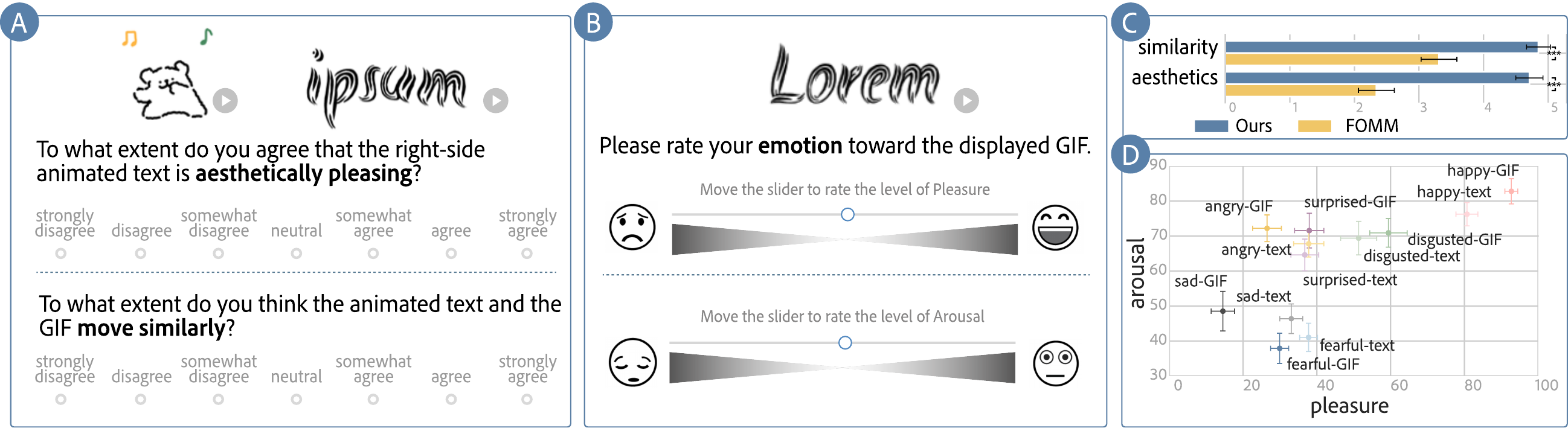}
  \caption{Example questions and results in two questionnaire studies (N=33). (A) Study I: Subjective ratings for the aesthetic property of the output kinetic typography and similarity between the target and source GIF. (B) Study II: Perception of emotions underlying kinetic typography. (C) The average ratings and standard errors of cases in Study I, where our approach outperforms the baseline FOMM in motion similarity and result aesthetics. (D) The average ratings and standard errors of the pairwise (pleasure, arousal) ratings for sample GIFs, where the driving GIF and corresponding kinetic typography posit in adjacent areas. \rev{Ratings of the same emotion are encoded with colors of a similar hue.}}
  \label{fig: crowd}
  \Description{Figure 10 contains four figures demonstrating the questionnaire design and results. (A) A snapshot of Questionnaire I. The top shows the anchor GIF and the corresponding kinetic typography. The bottom is a 7-point scale for rating motion similarity and aesthetic pleasure. (B) A snapshot of Questionnaire II. The top shows a GIF, and the bottom is a slider scale for rating pleasure and arousal. (C) The results of Questionnaire I show that our method is superior to FOMM in terms of aesthetic appeal and movement similarity. (D) The results of Questionnaire II show that the anchor GIF of a certain emotion and its corresponding generated kinetic typography are located closely on a two-dimensional coordinate system with pleasure and arousal as the x and y axes, respectively.}
\end{figure*}

A sample question is shown in \autoref{fig: crowd} A.
When designing the questionnaires, we tried to familiarize participants with simple and concrete questions.
For instance, we asked whether a GIF is ``aesthetically pleasing'' to align participants' appraisal of the aesthetic property to their feelings~\cite{AestheticPrinciples}\rev{.}
For each question, the animated text and the corresponding anchor GIF are displayed, and participants are asked to rate them on a 7-point Likert scale (0--strongly disagree to 6--strongly agree) for aesthetics and motion similarity, respectively. 
20 driving GIFs were randomly selected.
For each driving GIF, we set two questions, one for the baseline--rasterized kinetic typography generated with FOMM, and one for the experimental group--vectorized kinetic typography with our approach.
Therefore, a questionnaire consists of 40 questions.
Considering the influence of the font, participants are asked to rate the aesthetics of the static font before viewing the main body of the questionnaire.

\paragraph{Study II: Semantic Preservation}

The second study aimed to verify whether the learned animation can preserve the semantics of the driving GIF.
We tackled the problem from the perspective of emotion, which is an integral part of semantics.
Specifically, we focused on Ekman's six basic emotions~\cite{ekman1999basic}, \ie~sadness, happiness, fear, anger, surprise, and disgust.

\autoref{fig: crowd} B illustrates a sample question. We adopted the Affective Slider~\cite{betella2016affective} for participants to self-report their emotions, which consist of two dimensions: pleasure and arousal from the extent 1 to 100.
Pleasure means the degree of positivity or negativity of an individual's emotional state.
Arousal corresponds to the level of physiological activation or stimulation in an individual's emotional state.
For each basic emotion, we selected two GIFs as anchors \rev{according to the pre-defined labels in the dataset.}
A question comprises one kinetic typography, where we required participants to assess the perceived emotions.
Hence, there were 12 questions.


\subsubsection{Participants}
We recruited participants from a local university by posting advertisements on social media. Each participant is paid £3.5 for completing the questionnaire.
A total of 33 people signed up for the questionnaire study. 
Most participants were between 18--24 years old, with 15 females and 18 males.
In addition, all participants reported using emojis frequently in daily communication, where 
14 (42\%) reported to use emojis \textit{very often}.

\subsubsection{Result Analysis} 
\paragraph{Study I. Motion Transfer}
Participants spent an average of 10.4 minutes in completing the 20 questions (std=6.2, ranging from 3.5 to 28).
We deemed all the responses valid.
Seen from \autoref{fig: crowd} C, participants generally recognized the aesthetics and the similarity of movement between the driving GIFs and the animated text in our approach.
\rev{For our approach, t}he average score on the aesthetics was 4.71 (std=1.21), and the motion similarity was 4.85 (std=1.06), where both scores exceeded 4, \ie, somewhat agree, on the 7-point scale. \rev{For the baseline, the average aesthetic score was 2.34 (std=1.60) and the average similarity score was 3.31 (std=1.59).}
Compared to the baseline method, our approach obtained significantly higher scores (significance level $\alpha=0.001$, Student's t--test), which suggests that our method outperforms the FOMM in terms of motion transfer, making the generated animations more visually appealing and more similar to the source motion.
This finding echoes our ablation study on the vectorized text representation, where we identified certain glitches in the pixel-based methods.
Moreover, after adding the dynamic motion, the aesthetics of the text improves compared to static text with an average aesthetics score of 3.79, further validating the effectiveness of our method.

In summary, Study I verified that our model could achieve better motion transfer compared to the baseline. 
The improved aesthetics and motion similarity scores, along with the user comments, demonstrated the effectiveness and applicability of our method in generating visually appealing and motion-consistent animated text.
\begin{figure*}[h]
  \centering
  \includegraphics[width=\textwidth]{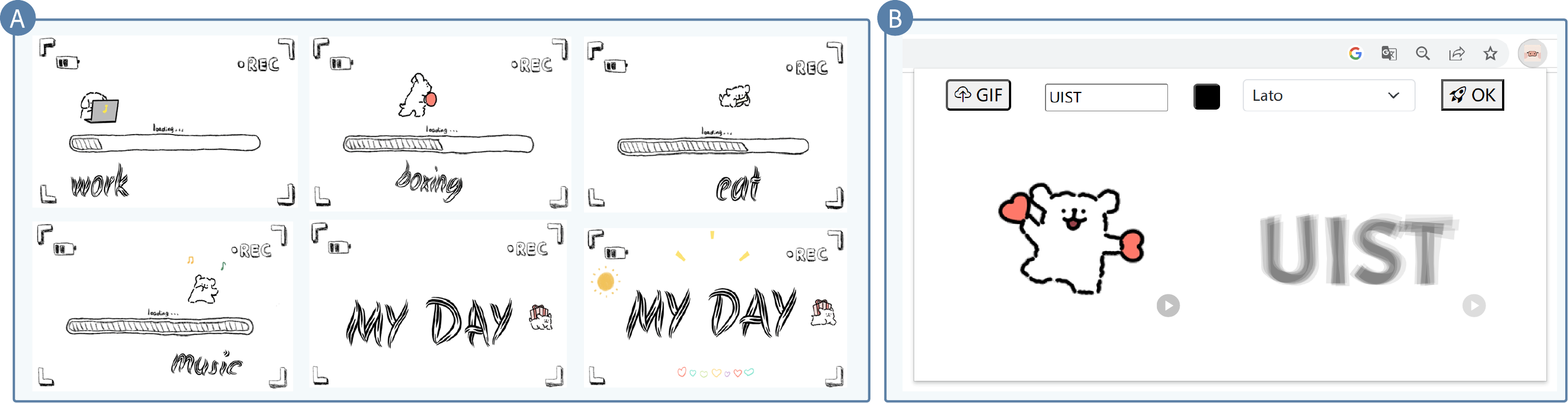}
  \caption{Demonstrations in the workshop. (A) A video opening featuring synchronous animation of textual descriptions and the cartoon character. (B) A browser plugin for the automatic generation of kinetic typography based on our approach.}
  \label{fig: video}
  \Description{Figure 11 shows two usage scenarios labeled (A) and (B). (A) Six frames of the video opening made with kinetic typography generated by our method. (B) The browser plugin interface with the top input area above and the bottom preview and output area.}
\end{figure*}
\paragraph{Study II. Emotion Preservation}
Participants spent an average of 8.5 minutes to complete the questionnaire, with a maximum of 30 minutes and a minimum of 2 minutes (std=8.9).
The results of study II are shown in \autoref{fig: crowd} B, where the scatterplot maps the average scores of the (pleasure, arousal) pairs.
The attached error bars indicate the standard error of each dimension with their lengths, where vertical for arousal and horizontal for pleasure.

Inspecting the diagram, we could see that the data points for the same emotion under both the driving GIF and Text are projected on adjacent areas, revealing that the expression of emotion has also been successfully transferred through the motion transfer. 
Furthermore, different emotion classes largely varied.
For example, there is a significant difference between happy and sad emotions, demonstrating the effectiveness of our method in preserving the emotional semantics of the anchor GIFs.
One could see that the emotions of disgust and anger are very close, and the demarcation is not obvious, with only the ordering of pleasure being altered. 
This suggested that the emotional expressions in these two emotions might be similar, making it harder for users to differentiate them clearly.
In addition, the variances in the arousal dimension were generally greater, possibly due to the users' perception of arousal being more ambiguous or subjective, leading to a wider range of responses.
In contrast, the arousal and pleasure scores for the text are more neutral (around 50 points). 
Although the emotional expression has been learned and transferred, it is not as strong as in the source motion pictures. 
It might result from the limitations of the method or the inherent difference between text and the figure-like domain in representing emotions.

In summary, results from Study II showed that our method could effectively preserve the emotional semantics when transferring animations from the driving GIF to a text. However, some emotions may not be as distinct as they are in the original anchor GIFs. And users' perceptions of arousal might be more ambiguous. \rev{While the questionnaire shows the success of emotion transfer on the particular typeface being used, more studies are needed to validate similar mechanisms for other fonts, as we did not eliminate the influence on emotion perception from the typeface.}


%% file: sections/08User.tex
\section{Workshop}
\label{sec:workshop}
We organized a workshop to evaluate the utility of our method.

\subsection{Demonstrations}
To elicit in-depth discussions in the workshop, we designed and implemented several demonstrations of potential application scenarios, including a video opening, an online messaging widget, and an emotional word cloud.

\paragraph{Video opening}

Vlogs (Video blogs) are becoming a prevalent form to share personal experiences creatively.
To make a vlog stand out, an engaging opening animation can help grab viewers' attention and set the tone for the rest of the video.
Leveraging results generated by our method, we created a vlog opening as an example of its practical application in daily content creation.
As shown in \autoref{fig: video} A, D1 is a vlog opening with the Puppy Maltese theme, showcasing different states of the character and introducing ``My Day'', which suggests the video topic. 
We envision that integrating animated text with consistent motions enhances both explanatory and entertaining values. 
Similarly, users can use our method to create and personalize animated text in their video creations across different themes and contents.

\paragraph{Browser Widget for Online Chatting}
Informed by the previous efforts in enhancing emotion communication in online messaging~\cite{Wang2004communicate, malik2009communicating, aoki2022emoballon}, we implemented a light-weighted Chrome extension to facilitate the real-time creation of kinetic typography (see \autoref{fig: video} B).
It has a simplified interface compared with Wakey-wakey, which features real-time generation and removes the human interaction module.
Users may upload an anchor GIF, type down the text, configure the color and font, and then directly obtain the generated kinetic typography in the GIF format.

\begin{figure}[t]
  \centering
  \includegraphics[width=\linewidth]{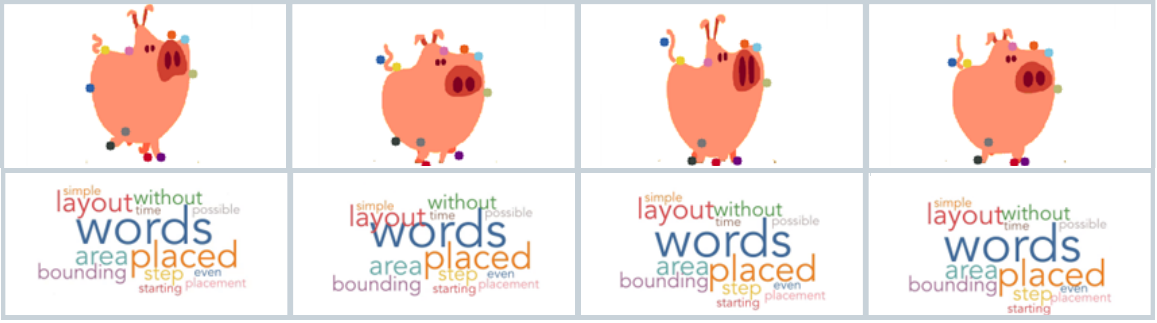}
  \caption{Application of animated word cloud that delivers a certain emotion with the animation.}
  \label{fig: wccase}
  \Description{In Figure 12, there are two rows of keyframes showing the anchor GIF and the corresponding animated word cloud.}
\end{figure}

\paragraph{Emotional Animated Word Cloud}
The word cloud is a common visualization technique to summarize text data, where the text size represents the word frequency.
\rev{Xie \ea~\cite{xie2023emordle} coined the word ``emordle'' representing animated word clouds that suggest underlying emotions.
Based on our approach, we generated an ``emordle'' by transferring the animated scheme of a bumpy cartoon pig from the MGif dataset.
Instead of using the parsed control points, we transformed the text anchors for each text element that constitute the word cloud. 
In other words, we replace the vector control points with the anchor points of each word in the proposed approach. Note that one word has one anchor point at its central position.}
The generated word cloud example is displayed in \autoref{fig: wccase}.

\subsection{Protocol}
The workshop proceeded in the following four stages.

\paragraph{Briefing}
The briefing session took 10 minutes, during which we introduced the background of our work and briefly explained our method. We then used a demo video to demonstrate the functions of the tool and illustrate the generation process.

\paragraph{Demonstration} 
We spent another 10 minutes displaying the video opening, the online messaging widget, and the animated word cloud to demonstrate potential application scenarios. We also introduced the installation and use of the plugin through a demo.

\paragraph{Self-creation}
We then allow users to freely use and explore the system and widget to create their kinetic typography for 20 minutes. 

\paragraph{Post-interview}
After trying \tool, participants are asked to complete a questionnaire with six questions on a 7-point Likert scale about its usability, including covering \textit{Practicality}, \textit{Customization}, \textit{Pleasure}, \textit{Efficiency}, and the \textit{Intuitiveness} for widget and interface respectively.
We also raised open-ended questions following a structured template in order to understand their perceptions of the authoring process as well as the generated kinetic typography.

\subsection{Participants}
Both designers and general users are invited to obtain feedback from different perspectives in the evaluation.
We recruited 20 participants through our personal network and advertisements on social media, with 7 females and 13 males.
There are 3 professional designers: P1 works on user experience design; P2 engages in self-media creation as a blogger and vlogger; P3 is a digital painter (P3).
The rest are graduate students majoring in data science at a local university (denoted as P4–P20 in ascending time order of their interviews).
Among all participants, 13 have seen kinetic typography before, such as in online memes, short videos, slides, and advertisements. 
Only two participants have experience in creating kinetic typography with other software (P2, P3). 

\subsection{Results}
Here we present both the quantitative and qualitative results.

\subsubsection{Observations.}
Nonetheless, when it comes to the time required for creation, there is substantial dissent, as our concurrency fell short, leading to extended completion times for some users.
In spite of this setback, the system garners favorable acknowledgment for its expressiveness, user-friendliness, intuitiveness, and real-world applicability.
To elevate the overall user experience, refinements in creation duration and concurrency should be considered.
\rev{As for the usage of $\alpha$ or $e$, most users were generally satisfied with the empirical default value. However, two users with a design background occasionally fine-tuned this parameter to derive better results.
In terms of the manual adjustments on control points, users without a design background barely adjusted the control points. Three users with design backgrounds occasionally make manual adjustments, about 8/14 frames, with an average duration of 2.1/5.8 min for one kinetic typography. Users tended to adjust key points in GIFs when there were noticeable detection errors. Otherwise, they normally increased $\alpha$ to mitigate unexpected deformations, though less motion preserving.}

\subsubsection{Usability Ratings.}

\begin{figure}[t]
  \centering
  \includegraphics[width=\linewidth]{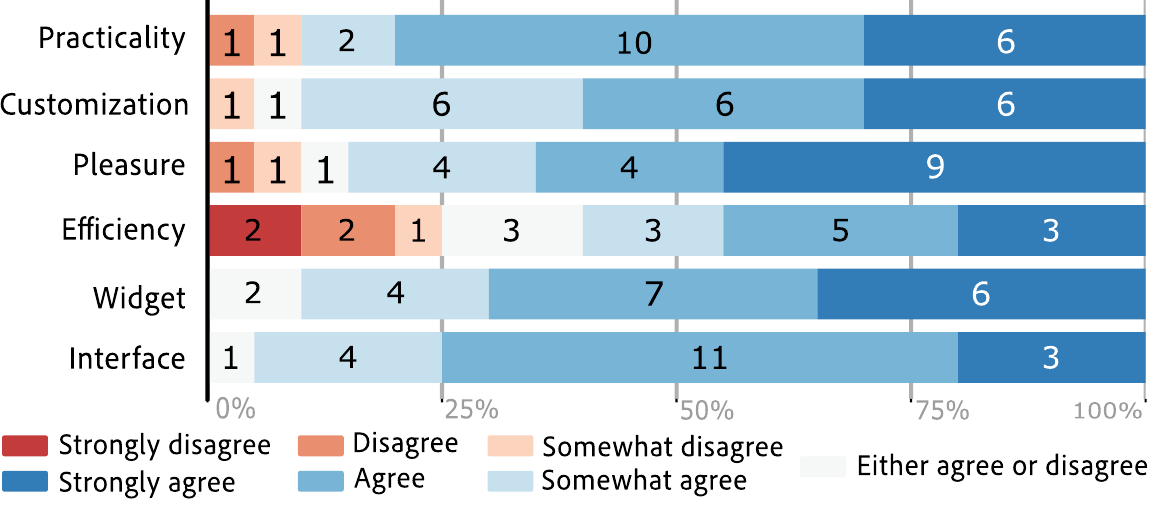}
  \caption{Quantitative Results of the usefulness of our method. We measured the \textit{practicality}, \textit{customization}, \textit{pleasure}, \textit{efficiency} and the \textit{intuitiveness} for the browser widget and authoring interface respectively.}
  \label{fig: quant_res}
  \Description{Figure 13 shows the overall chart is a stacked bar chart with seven categories ranging from "strongly disagree" to "strongly agree". Vertically, it represents "practicality", "customization", "pleasure", "efficiency", "widget", and "interface". The x-axis is marked with percentages and the specific statistics are indicated on the bars. Our approach has been recognized for its usefulness.}
\end{figure}

\autoref{fig: quant_res} shows the distribution of users' subjective ratings of the six usability questions.
More than half of users concurred that the tool was intuitive, tailored, and delivers an enjoyable experience during usage.
 To be more specific, 45\% users strongly agreed that our work is pleasant and interesting.
 They valued the innovative animated text design, the straightforward comprehension of emotions portrayed, and the uncomplicated tool configurations for selecting animation approaches.
 They also observed that blending textual meanings with animations facilitated a beneficial expression of the content.

\subsubsection{User Feedback.}
We summarize the following insights and implications for future improvements from the users' feedback.

$\diamond$ \underline{On mixed-initiative authoring}. Participants expressed their agreement with the balance we stroke between user involvement and automatic generation. P3 said: ``{\it supporting quick automatic generation while also offering optional customization and improvement from users}''. Participants were optimistic about the mixed-initiative way, as ``{\it intelligence improves efficiency while users enhance quality and creativity, as the imagination of users cannot be dismissed}'' (P10).

$\diamond$ \underline{On personalization.}
19 participants (except P11) valued customization highly and believed it helps incorporate their own ideas and shape a unique personal style, while three designers showed an ``{\it innate aversion to preconceived solutions}'' (P1). Among them, 17 believed that customization is also crucial in the context of animated text, which helps to ``{\it express opinions and feelings more effectively and precisely}'' (P10), as well as ``{\it making the creations more specific and memorable}'' (P18).

$\diamond$ \underline{On motion transfer.}
Users expressed their agreement with the novelty, interest, and inspiration of our approach.
Seven participants mentioned ``novel'', where P9 commented ``{\it it is pleasantly surprising}''.
Four participants mentioned ``interesting'', and 3 participants found it inspiring.
P2 found {``\it the generated results inspiring and insightful for my design progress''}.
P1 said: ``{\it I appreciate the smart use of motion transfer, as the interpretation of motion is subjective, but you use memes as an intermediate medium whose ability to convey emotions is validated through wide applications, making the generated results meet subjective expectations. In addition, using text as a vector container for transformation, where the container can be liquefied, allows for slight deformation, thus in support of more subtle emotional expression.}''

$\diamond$ \underline{On application scenarios}.
Participants brainstormed multiple application scenarios in their personal life with kinetic typography, including online chatting, social media post, website banners, presentations, subtitle enhancement, and E-invitation. 
\rev{P13 also imagined that in order to attract interest, animated text may be used as a teaching instrument for introducing words to little children.}
P1 identified some challenges in the application of the animated text. ``{\it When used alone, the interpretation by users can be ambiguous. There are challenges in accessibility, readability, as well as efficiency of perception and recognition}''.
He suggested that we can ``{\it emphasize the combination with animated images, which can enhance contextual effects and create a synergistic interaction greater than the sum of the separate parts}''.

$\diamond$ \underline{On future improvements}. 
First, Participants suggested a possible enhancement in the guidance of interactions, preferably by introducing recommendations for interactive operations. 
The operations may be clearer ``{\it with the help of some icons and text}'' (P16), and ``{\it the generation efficiency can be improved by recommending interactive behaviors. In addition, it would be very helpful to support automatic modifications of other frames after modifying one frame in key point correction and glyph refinement}'' (P2).
Besides, integrating external resources may enrich the generated results.
P11 suggested ``{\it incorporating language models to generate using instructions enhances its convenience}''.
P3 commented that ``{\it integrating meme and artistic font libraries helps generate richer and more artistic results}''.

%% file: sections/09Discussion.tex
\section{Discussion}
\label{sec:discussion}
\rev{We summarize the implications of our investigation, reflect on our limitations, and discuss promising directions for future research.}

\subsection{Implication}
\rev{
\paragraph{Create animated effects with model-free motion transfer.}
We contribute a novel approach in the emerging area of AI-generated content to design motion graphics using prevalent cross-domain GIFs as references.
Participants in the workshop acknowledge the ease of guiding animation generation with reference GIFs.
Despite various properties to coordinate in animation design, the underlying workflow of \tool{} helps users to author in a top-down manner instead of tweaking every details.
Beyond template engines, model-free motion transfer helps create more diversified effects.

\paragraph{Support human-AI collaboration with interpretable features.}
According to the user feedback in the workshop, the extracted key point helps them understand the causes of misalignment.
While most of our participants are unaware of the internal mechanism, they are able to correct the flaws introduced by the black-box model and intervene in the generation process to produce more desirable results.
In designing AI-empowered authoring tools, interpretable features are practical entry points for humans to inject requirements into the content creation process.

\paragraph{Consider design requirements of users at different levels.}
\tool{} supports both one-off generation and fine-grain control.
On the one hand, there are default values underlying algorithms to cater to the fast-generation need of causal users.
On the other hand, configurable parameters and the vector representation of generated results are also exposed for further adjustment.
When developing authoring support for prevailing artifacts, such as kinetic typography or data visualization, it is important to consider the requirements of different user profiles to make the authoring tool more useful.
}

\subsection{Limitation}

\paragraph{Deformation stability.} 
When the motion amplitude of the character in the driving GIF is too large, excessive deformation may occur in the glyph, especially for fonts with only a few control points.
Severe deformation can result in distorted glyphs with discontinued outlines and low legibility.
This may be addressed by expanding the number of control points along the predefined glyph outline~\cite{iluz2023word}, using the triangulated mesh representation of glyphs~\cite{desbrun1999implicit}, \rev{and introducing global penalty in the loss function}.
\rev{In addition, as discussed in \autoref{sec:generalizability}, the automatic pipeline may fail for over-complicated GIF styles or typefaces, which requires more generalized models in motion trajectory extraction.}  


\paragraph{\rev{Motion semantic perseverance.}} 
\rev{
As with other cross-domain motion transfer problems, when generated result may not preserve the original semantics, as text generally lacks comparable internal structures with GIFs, where the length of a text strongly influences the success.
With our approach, a ``goodbye'' may crawl like a snake but hardly a giraffe swinging its neck.
In addition, our objective function prioritizes the overall deformation of the exterior and may neglect the local and independent deformation of the interior.
For instance, the generated result may learn the waving gesture but miss the delicate eye blinking.}

\paragraph{\rev{Animation expressiveness.}}
We leverage motion transfer on text control points to deform the shape of text elements and mimic the general animated effects in the driving GIF.
However, in addition to learning the shape-deformation patterns, kinetic typography also concerns other properties~\cite{xie2023emordle}.
Future works may explore incorporating visual properties like colors and designing an integrated environment for a more flexible authoring experience.

\subsection{Future Work}


\rev{This work demonstrates a novice-friendly approach to creating text animation through cross-domain motion transfer.
While we focus on texts, it is also interesting to explore arbitrary anthropomorphized shapes, such as the dancing mushrooms in Disney's \textit{Fantasia}~\cite{fantasia} or sketches of monsters~\cite{su2018live,smith2023tog}.
Unlike human postures constrained by bones and flesh, motion graphics enjoy higher flexibility for exaggerating effects, which share similarities with text.
We note that the outline optimization should shift the focus from maintaining text legibility to shape semantics. 
Recent advances in large language-vision models like Stable Diffusion~\cite{rombach2022high} may help to regularize undesired artifacts.
In addition, animated data storytelling (\eg,~\cite{wang2021infomotion,Shu21Gif}) remains an exciting avenue for integrating expressive motions on visual marks.
As illustrated in the case of an animated word cloud (see \autoref{fig: wccase}), the positions of individual visual marks in a visualization can be regarded as the control points of a text.
Using our approach, it is possible to transfer motion into visualizations, which may extend existing visual vocabularies and further facilitate the comprehension of abstract data and the expression of emotions~\cite{lan21kineticharts, xie2023emordle}. }

%% file: sections/10Conclusion.tex
\section{Conclusion}
In this study, we explore the opportunity to create expressive kinetic typography based on a driving GIF with character motions.
Based on the unique characteristics of text, we propose a framework that adapts existing motion transfer models to the vector domain.
Specifically, we animate text based on their control points predefined in font specification.
To mimic the motion while maintaining fair legibility, the by-frame positions of each control point are regularized by the extracted motion key point in the driving GIF and neighboring control points.
We also introduce an interaction module that allows human-computer collaboration, where humans can steer the intermediate results and guide the generation of kinetic typography.
Based on the framework, we developed a mixed-initiative authoring tool and a browser widget featuring automatic generation.
A questionnaire study (N=33) initially validated the effectiveness of our approach.
Participants generally recognized that the results were animated in a desirable manner, both aesthetically pleasing and semantically resonant.
Moreover, we evaluated the novel transfer-based kinetic typography tools by organizing a workshop (N=20) with both general users and professional designers.
People were positive about the interactive system and showed interest in employing our tools for various scenarios.

%% file: sections/acks.tex

\begin{acks}
This research was supported by the Natural Science Foundation of China (NSFC No.62202105), Shanghai Municipal Science and Technology (No. 21ZR1403300 and No. 21YF1402900), and Hong Kong Research Grants Council General Research Fund 16210722.
We thank the anonymous reviewers, participants in the user studies, Zhan Wang, Ziyue Lin, Dr. Xinhuan Shu, Dr. Jiaxiong Hu, and Prof. Zhenjie Zhao for valuable feedback.
\end{acks}

%% file: sections/appendix.tex
\appendix

%% file: main.bbl

\begin{thebibliography}{70}


\ifx \showCODEN    \undefined \def \showCODEN     #1{\unskip}     \fi
\ifx \showDOI      \undefined \def \showDOI       #1{#1}\fi
\ifx \showISBNx    \undefined \def \showISBNx     #1{\unskip}     \fi
\ifx \showISBNxiii \undefined \def \showISBNxiii  #1{\unskip}     \fi
\ifx \showISSN     \undefined \def \showISSN      #1{\unskip}     \fi
\ifx \showLCCN     \undefined \def \showLCCN      #1{\unskip}     \fi
\ifx \shownote     \undefined \def \shownote      #1{#1}          \fi
\ifx \showarticletitle \undefined \def \showarticletitle #1{#1}   \fi
\ifx \showURL      \undefined \def \showURL       {\relax}        \fi
\providecommand\bibfield[2]{#2}
\providecommand\bibinfo[2]{#2}
\providecommand\natexlab[1]{#1}
\providecommand\showeprint[2][]{arXiv:#2}

\bibitem[{Adele}(2012)]%
        {skyfall}
\bibfield{author}{\bibinfo{person}{{Adele}}.} \bibinfo{year}{2012}\natexlab{}.
\newblock \bibinfo{booktitle}{\emph{Adele - Skyfall (Official Lyric Video)}}.
\newblock
\urldef\tempurl%
\url{https://youtu.be/DeumyOzKqgI}
\showURL{%
Retrieved May 30, 2023 from \tempurl}
\newblock
\shownote{1:37--1:53}.


\bibitem[{Adobe Inc.}(2023)]%
        {aftereffects}
\bibfield{author}{\bibinfo{person}{{Adobe Inc.}}}
  \bibinfo{year}{2023}\natexlab{}.
\newblock \bibinfo{booktitle}{\emph{After Effects}}.
\newblock
\urldef\tempurl%
\url{https://www.adobe.com/products/aftereffects.html}
\showURL{%
Retrieved Mar 20, 2023 from \tempurl}


\bibitem[Aoki et~al\mbox{.}(2022)]%
        {aoki2022emoballon}
\bibfield{author}{\bibinfo{person}{Toshiki Aoki}, \bibinfo{person}{Rintaro
  Chujo}, \bibinfo{person}{Katsufumi Matsui}, \bibinfo{person}{Saemi Choi},
  {and} \bibinfo{person}{Ari Hautasaari}.} \bibinfo{year}{2022}\natexlab{}.
\newblock \showarticletitle{EmoBalloon--Conveying Emotional Arousal in Text
  Chats with Speech Balloons}. In \bibinfo{booktitle}{\emph{Proceedings of the
  ACM Conference on Human Factors in Computing Systems (CHI)}}.
  \bibinfo{publisher}{ACM}, \bibinfo{address}{New York, NY, USA}, Article
  \bibinfo{articleno}{527}, \bibinfo{numpages}{16}~pages.
\newblock
\urldef\tempurl%
\url{https://doi.org/10.1145/3491102.3501920}
\showDOI{\tempurl}


\bibitem[{Apple Inc.}(2023)]%
        {motion}
\bibfield{author}{\bibinfo{person}{{Apple Inc.}}}
  \bibinfo{year}{2023}\natexlab{}.
\newblock \bibinfo{booktitle}{\emph{Motion}}.
\newblock
\urldef\tempurl%
\url{https://www.apple.com/final-cut-pro/motion/}
\showURL{%
Retrieved Mar 20, 2023 from \tempurl}


\bibitem[Arora et~al\mbox{.}(2019)]%
        {arora2019magicalhands}
\bibfield{author}{\bibinfo{person}{Rahul Arora}, \bibinfo{person}{Rubaiat~Habib
  Kazi}, \bibinfo{person}{Danny~M Kaufman}, \bibinfo{person}{Wilmot Li}, {and}
  \bibinfo{person}{Karan Singh}.} \bibinfo{year}{2019}\natexlab{}.
\newblock \showarticletitle{Magicalhands: Mid-Air Hand Gestures for Animating
  in VR}. In \bibinfo{booktitle}{\emph{Proceedings of the ACM Annual Symposium
  on User Interface Software and Technology (UIST)}}. \bibinfo{publisher}{ACM},
  \bibinfo{address}{New York, NY, USA}, \bibinfo{pages}{463--477}.
\newblock
\urldef\tempurl%
\url{https://doi.org/10.1145/3332165.3347942}
\showDOI{\tempurl}


\bibitem[Betella and Verschure(2016)]%
        {betella2016affective}
\bibfield{author}{\bibinfo{person}{Alberto Betella} {and}
  \bibinfo{person}{Paul~FMJ Verschure}.} \bibinfo{year}{2016}\natexlab{}.
\newblock \showarticletitle{The Affective Slider: A Digital Self-assessment
  Scale for the Measurement of Human Emotions}.
\newblock \bibinfo{journal}{\emph{PloS one}} \bibinfo{volume}{11},
  \bibinfo{number}{2}, Article \bibinfo{articleno}{e0148037}
  (\bibinfo{year}{2016}), \bibinfo{numpages}{11}~pages.
\newblock
\urldef\tempurl%
\url{https://doi.org/10.1371/journal.pone.0148037}
\showDOI{\tempurl}


\bibitem[Borzyskowski(2004)]%
        {borzyskowski2004animated}
\bibfield{author}{\bibinfo{person}{George Borzyskowski}.}
  \bibinfo{year}{2004}\natexlab{}.
\newblock \showarticletitle{Animated Text: More than Meets the Eye?}. In
  \bibinfo{booktitle}{\emph{Proceedings of the ASCILITE Conference}}.
  \bibinfo{pages}{141--144}.
\newblock
\urldef\tempurl%
\url{https://www.ascilite.org/conferences/perth04/procs/pdf/borzyskowski.pdf}
\showURL{%
\tempurl}


\bibitem[Chan et~al\mbox{.}(2019)]%
        {chan2019everybody}
\bibfield{author}{\bibinfo{person}{Caroline Chan}, \bibinfo{person}{Shiry
  Ginosar}, \bibinfo{person}{Tinghui Zhou}, {and} \bibinfo{person}{Alexei~A
  Efros}.} \bibinfo{year}{2019}\natexlab{}.
\newblock \showarticletitle{Everybody Dance Now}. In
  \bibinfo{booktitle}{\emph{Proceedings of the IEEE/CVF International
  Conference on Computer Vision (ICCV)}}. \bibinfo{publisher}{IEEE},
  \bibinfo{address}{Piscataway, NJ, USA}, \bibinfo{pages}{5933--5942}.
\newblock
\urldef\tempurl%
\url{https://doi.org/10.1109/ICCV.2019.00603}
\showDOI{\tempurl}


\bibitem[Chang and Ungar(1993)]%
        {chang1993animation}
\bibfield{author}{\bibinfo{person}{Bay-Wei Chang} {and} \bibinfo{person}{David
  Ungar}.} \bibinfo{year}{1993}\natexlab{}.
\newblock \showarticletitle{Animation: From Cartoons to the User Interface}. In
  \bibinfo{booktitle}{\emph{Proceedings of the ACM Annual Symposium on User
  Interface Software and Technology (UIST)}}. \bibinfo{publisher}{ACM},
  \bibinfo{address}{New York, NY, USA}, \bibinfo{pages}{45--55}.
\newblock
\urldef\tempurl%
\url{https://doi.org/10.1145/168642.168647}
\showDOI{\tempurl}


\bibitem[Chuang et~al\mbox{.}(2005)]%
        {chuang2005animating}
\bibfield{author}{\bibinfo{person}{Yung-Yu Chuang}, \bibinfo{person}{Dan~B
  Goldman}, \bibinfo{person}{Ke~Colin Zheng}, \bibinfo{person}{Brian Curless},
  \bibinfo{person}{David~H. Salesin}, {and} \bibinfo{person}{Richard
  Szeliski}.} \bibinfo{year}{2005}\natexlab{}.
\newblock \showarticletitle{Animating Pictures with Stochastic Motion
  Textures}.
\newblock \bibinfo{journal}{\emph{ACM Transactions on Graphics}}
  \bibinfo{volume}{24}, \bibinfo{number}{3} (\bibinfo{year}{2005}),
  \bibinfo{pages}{853--860}.
\newblock
\urldef\tempurl%
\url{https://doi.org/10.1145/1073204.1073273}
\showDOI{\tempurl}


\bibitem[Conolly(2003)]%
        {AestheticPrinciples}
\bibfield{author}{\bibinfo{person}{O. Conolly}.}
  \bibinfo{year}{2003}\natexlab{}.
\newblock \showarticletitle{Aesthetic Principles}.
\newblock \bibinfo{journal}{\emph{The British Journal of Aesthetics}}
  \bibinfo{volume}{43} (\bibinfo{date}{04} \bibinfo{year}{2003}),
  \bibinfo{pages}{114--125}.
\newblock
\urldef\tempurl%
\url{https://doi.org/10.1093/bjaesthetics/43.2.114}
\showDOI{\tempurl}


\bibitem[Desbrun et~al\mbox{.}(1999)]%
        {desbrun1999implicit}
\bibfield{author}{\bibinfo{person}{Mathieu Desbrun}, \bibinfo{person}{Mark
  Meyer}, \bibinfo{person}{Peter Schr{\"o}der}, {and} \bibinfo{person}{Alan~H
  Barr}.} \bibinfo{year}{1999}\natexlab{}.
\newblock \showarticletitle{Implicit Fairing of Irregular Meshes Using
  Diffusion and Curvature Flow}. In \bibinfo{booktitle}{\emph{Proceedings of
  the Annual Conference on Computer Graphics and Interactive Techniques
  (SIGGRAPH)}}. \bibinfo{publisher}{ACM}, \bibinfo{address}{New York, NY, USA},
  \bibinfo{pages}{317--324}.
\newblock
\urldef\tempurl%
\url{https://doi.org/10.1145/311535.311576}
\showDOI{\tempurl}


\bibitem[Devendorf and Ryokai(2013)]%
        {laura2013anytype}
\bibfield{author}{\bibinfo{person}{Laura Devendorf} {and}
  \bibinfo{person}{Kimiko Ryokai}.} \bibinfo{year}{2013}\natexlab{}.
\newblock \showarticletitle{AnyType: Provoking Reflection and Exploration with
  Aesthetic Interaction}. In \bibinfo{booktitle}{\emph{Proceedings of the ACM
  Conference on Human Factors in Computing Systems (CHI)}}.
  \bibinfo{publisher}{ACM}, \bibinfo{address}{New York, NY, USA},
  \bibinfo{pages}{1041--1050}.
\newblock
\urldef\tempurl%
\url{https://doi.org/10.1145/2470654.2466133}
\showDOI{\tempurl}


\bibitem[Dvoro\v{z}\v{n}\'{a}k et~al\mbox{.}(2017)]%
        {dvoroznak2017example}
\bibfield{author}{\bibinfo{person}{Marek Dvoro\v{z}\v{n}\'{a}k},
  \bibinfo{person}{Pierre B\'{e}nard}, \bibinfo{person}{Pascal Barla},
  \bibinfo{person}{Oliver Wang}, {and} \bibinfo{person}{Daniel S\'{y}kora}.}
  \bibinfo{year}{2017}\natexlab{}.
\newblock \showarticletitle{Example-Based Expressive Animation of 2D Rigid
  Bodies}.
\newblock \bibinfo{journal}{\emph{ACM Transactions on Graphics}}
  \bibinfo{volume}{36}, \bibinfo{number}{4}, Article \bibinfo{articleno}{127}
  (\bibinfo{year}{2017}), \bibinfo{numpages}{10}~pages.
\newblock
\urldef\tempurl%
\url{https://doi.org/10.1145/3072959.3073611}
\showDOI{\tempurl}


\bibitem[Ekman(1999)]%
        {ekman1999basic}
\bibfield{author}{\bibinfo{person}{Paul Ekman}.}
  \bibinfo{year}{1999}\natexlab{}.
\newblock \showarticletitle{Basic Emotions}.
\newblock \bibinfo{journal}{\emph{Handbook of Cognition and Emotion}}
  \bibinfo{volume}{98}, \bibinfo{number}{45--60} (\bibinfo{year}{1999}),
  \bibinfo{pages}{16}.
\newblock
\urldef\tempurl%
\url{https://doi.org/10.1002/0470013494.ch3}
\showDOI{\tempurl}


\bibitem[Ford et~al\mbox{.}(1997)]%
        {shannon1998kinetic}
\bibfield{author}{\bibinfo{person}{Shannon Ford}, \bibinfo{person}{Jodi
  Forlizzi}, {and} \bibinfo{person}{Suguru Ishizaki}.}
  \bibinfo{year}{1997}\natexlab{}.
\newblock \showarticletitle{Kinetic Typography: Issues in Time-Based
  Presentation of Text}. In \bibinfo{booktitle}{\emph{Extended Abstracts on
  Human Factors in Computing Systems (CHIEA)}}. \bibinfo{publisher}{ACM},
  \bibinfo{address}{New York, NY, USA}, \bibinfo{pages}{269--270}.
\newblock
\urldef\tempurl%
\url{https://doi.org/10.1145/1120212.1120387}
\showDOI{\tempurl}


\bibitem[Forlizzi et~al\mbox{.}(2003)]%
        {forlizzi2003kinedit}
\bibfield{author}{\bibinfo{person}{Jodi Forlizzi}, \bibinfo{person}{Johnny
  Lee}, {and} \bibinfo{person}{Scott Hudson}.} \bibinfo{year}{2003}\natexlab{}.
\newblock \showarticletitle{The Kinedit System: Affective Messages Using
  Dynamic Texts}. In \bibinfo{booktitle}{\emph{Proceedings of the ACM
  Conference on Human Factors in Computing Systems (CHI)}}.
  \bibinfo{publisher}{ACM}, \bibinfo{address}{New York, NY, USA},
  \bibinfo{pages}{377--384}.
\newblock
\urldef\tempurl%
\url{https://doi.org/10.1145/642611.642677}
\showDOI{\tempurl}


\bibitem[Gatys et~al\mbox{.}(2016)]%
        {gatys2016image}
\bibfield{author}{\bibinfo{person}{Leon~A Gatys}, \bibinfo{person}{Alexander~S
  Ecker}, {and} \bibinfo{person}{Matthias Bethge}.}
  \bibinfo{year}{2016}\natexlab{}.
\newblock \showarticletitle{Image Style Transfer Using Convolutional Neural
  Networks}. In \bibinfo{booktitle}{\emph{Proceedings of the IEEE Conference on
  Computer Vision and Pattern Recognition (CVPR)}}. \bibinfo{publisher}{IEEE},
  \bibinfo{address}{Piscataway, NJ, USA}, \bibinfo{pages}{2414--2423}.
\newblock
\urldef\tempurl%
\url{https://doi.org/10.1109/CVPR.2016.265}
\showDOI{\tempurl}


\bibitem[Gaylord et~al\mbox{.}(2015a)]%
        {gaylord2015body}
\bibfield{author}{\bibinfo{person}{Weston Gaylord}, \bibinfo{person}{Vivian
  Hare}, {and} \bibinfo{person}{Ashley Ngu}.} \bibinfo{year}{2015}\natexlab{a}.
\newblock \showarticletitle{Adding Body Motion and Intonation to Instant
  Messaging with Animation}. In \bibinfo{booktitle}{\emph{Adjunct Proceedings
  of the ACM Symposium on User Interface Software and Technology}}.
  \bibinfo{publisher}{ACM}, \bibinfo{address}{New York, NY, USA},
  \bibinfo{pages}{105--106}.
\newblock
\urldef\tempurl%
\url{https://doi.org/10.1145/2815585.2815741}
\showDOI{\tempurl}


\bibitem[Gaylord et~al\mbox{.}(2015b)]%
        {gaylord2015atim}
\bibfield{author}{\bibinfo{person}{Weston Gaylord}, \bibinfo{person}{Vivian
  Hare}, {and} \bibinfo{person}{Ashley Ngu}.} \bibinfo{year}{2015}\natexlab{b}.
\newblock \showarticletitle{Adding Body Motion and Intonation to Instant
  Messaging with Animation}. In \bibinfo{booktitle}{\emph{Adjunct Proceedings
  of the ACM Symposium on User Interface Software \& Technology (UIST
  Adjunct)}}. \bibinfo{publisher}{ACM}, \bibinfo{address}{New York, NY, USA},
  \bibinfo{pages}{105--106}.
\newblock
\urldef\tempurl%
\url{https://doi.org/10.1145/2815585.2815741}
\showDOI{\tempurl}


\bibitem[Halperin et~al\mbox{.}(2021)]%
        {halperin2021endless}
\bibfield{author}{\bibinfo{person}{Tavi Halperin}, \bibinfo{person}{Hanit
  Hakim}, \bibinfo{person}{Orestis Vantzos}, \bibinfo{person}{Gershon Hochman},
  \bibinfo{person}{Netai Benaim}, \bibinfo{person}{Lior Sassy},
  \bibinfo{person}{Michael Kupchik}, \bibinfo{person}{Ofir Bibi}, {and}
  \bibinfo{person}{Ohad Fried}.} \bibinfo{year}{2021}\natexlab{}.
\newblock \showarticletitle{Endless Loops: Detecting and Animating Periodic
  Patterns in Still Images}.
\newblock \bibinfo{journal}{\emph{ACM Transactions on Graphics}}
  \bibinfo{volume}{40}, \bibinfo{number}{4}, Article \bibinfo{articleno}{142}
  (\bibinfo{year}{2021}), \bibinfo{numpages}{12}~pages.
\newblock
\urldef\tempurl%
\url{https://doi.org/10.1145/3450626.3459935}
\showDOI{\tempurl}


\bibitem[Hong et~al\mbox{.}(2022)]%
        {hong2022depth}
\bibfield{author}{\bibinfo{person}{Fa-Ting Hong}, \bibinfo{person}{Longhao
  Zhang}, \bibinfo{person}{Li Shen}, {and} \bibinfo{person}{Dan Xu}.}
  \bibinfo{year}{2022}\natexlab{}.
\newblock \showarticletitle{Depth-Aware Generative Adversarial Network for
  Talking Head Video Generation}. In \bibinfo{booktitle}{\emph{Proceedings of
  the IEEE/CVF Conference on Computer Vision and Pattern Recognition (CVPR)}}.
  \bibinfo{publisher}{IEEE}, \bibinfo{address}{Piscataway, NJ, USA},
  \bibinfo{pages}{3397--3406}.
\newblock
\urldef\tempurl%
\url{https://doi.org/10.1109/CVPR52688.2022.00339}
\showDOI{\tempurl}


\bibitem[Iluz et~al\mbox{.}(2023)]%
        {iluz2023word}
\bibfield{author}{\bibinfo{person}{Shir Iluz}, \bibinfo{person}{Yael Vinker},
  \bibinfo{person}{Amir Hertz}, \bibinfo{person}{Daniel Berio},
  \bibinfo{person}{Daniel Cohen-Or}, {and} \bibinfo{person}{Ariel Shamir}.}
  \bibinfo{year}{2023}\natexlab{}.
\newblock \bibinfo{title}{Word-As-Image for Semantic Typography}.
\newblock
\newblock
\showeprint{2303.01818}
\newblock
\shownote{Accepted in ACM SIGGRAPH 2023}.


\bibitem[Kato et~al\mbox{.}(2015)]%
        {kato2015textalive}
\bibfield{author}{\bibinfo{person}{Jun Kato}, \bibinfo{person}{Tomoyasu
  Nakano}, {and} \bibinfo{person}{Masataka Goto}.}
  \bibinfo{year}{2015}\natexlab{}.
\newblock \showarticletitle{TextAlive: Integrated Design Environment for
  Kinetic Typography}. In \bibinfo{booktitle}{\emph{Proceedings of the ACM
  Conference on Human Factors in Computing Systems (CHI)}}.
  \bibinfo{publisher}{ACM}, \bibinfo{address}{New York, NY, USA},
  \bibinfo{pages}{3403--3412}.
\newblock
\showISBNx{9781450331456}
\urldef\tempurl%
\url{https://doi.org/10.1145/2702123.2702140}
\showDOI{\tempurl}


\bibitem[Kazi et~al\mbox{.}(2014)]%
        {kazi2014draco}
\bibfield{author}{\bibinfo{person}{Rubaiat~Habib Kazi}, \bibinfo{person}{Fanny
  Chevalier}, \bibinfo{person}{Tovi Grossman}, \bibinfo{person}{Shengdong
  Zhao}, {and} \bibinfo{person}{George Fitzmaurice}.}
  \bibinfo{year}{2014}\natexlab{}.
\newblock \showarticletitle{Draco: Bringing Life to Illustrations with Kinetic
  Textures}. In \bibinfo{booktitle}{\emph{Proceedings of the ACM Conference on
  Human Factors in Computing Systems (CHI)}}. \bibinfo{publisher}{ACM},
  \bibinfo{address}{New York, NY, USA}, \bibinfo{pages}{351--360}.
\newblock
\urldef\tempurl%
\url{https://doi.org/10.1145/2556288.2556987}
\showDOI{\tempurl}


\bibitem[Kazi et~al\mbox{.}(2016)]%
        {kazi2016motionamplifiers}
\bibfield{author}{\bibinfo{person}{Rubaiat~Habib Kazi}, \bibinfo{person}{Tovi
  Grossman}, \bibinfo{person}{Nobuyuki Umetani}, {and} \bibinfo{person}{George
  Fitzmaurice}.} \bibinfo{year}{2016}\natexlab{}.
\newblock \showarticletitle{Motion Amplifiers: Sketching Dynamic Illustrations
  Using the Principles of 2D Animation}. In
  \bibinfo{booktitle}{\emph{Proceedings of the ACM Conference on Human Factors
  in Computing Systems (CHI)}}. \bibinfo{publisher}{ACM}, \bibinfo{address}{New
  York, NY, USA}, \bibinfo{pages}{4599--4609}.
\newblock
\urldef\tempurl%
\url{https://doi.org/10.1145/2858036.2858386}
\showDOI{\tempurl}


\bibitem[Kim et~al\mbox{.}(2016)]%
        {kim2016yo}
\bibfield{author}{\bibinfo{person}{Minhwan Kim}, \bibinfo{person}{Kyungah
  Choi}, {and} \bibinfo{person}{Hyeon-Jeong Suk}.}
  \bibinfo{year}{2016}\natexlab{}.
\newblock \showarticletitle{Yo! Enriching Emotional Quality of Single-Button
  Messengers through Kinetic Typography}. In
  \bibinfo{booktitle}{\emph{Proceedings of the ACM Conference on Designing
  Interactive Systems (DIS)}}. \bibinfo{publisher}{ACM}, \bibinfo{address}{New
  York, NY, USA}, \bibinfo{pages}{276--280}.
\newblock
\urldef\tempurl%
\url{https://doi.org/10.1145/2901790.2901835}
\showDOI{\tempurl}


\bibitem[Lai et~al\mbox{.}(2016)]%
        {lai2016data}
\bibfield{author}{\bibinfo{person}{Yu-Chi Lai}, \bibinfo{person}{Bo-An Chen},
  \bibinfo{person}{Kuo-Wei Chen}, \bibinfo{person}{Wei-Lin Si},
  \bibinfo{person}{Chih-Yuan Yao}, {and} \bibinfo{person}{Eugene Zhang}.}
  \bibinfo{year}{2016}\natexlab{}.
\newblock \showarticletitle{Data-driven NPR Illustrations of Natural Flows in
  Chinese Painting}.
\newblock \bibinfo{journal}{\emph{IEEE Transactions on Visualization and
  Computer Graphics}} \bibinfo{volume}{23}, \bibinfo{number}{12}
  (\bibinfo{year}{2016}), \bibinfo{pages}{2535--2549}.
\newblock
\urldef\tempurl%
\url{https://doi.org/10.1109/TVCG.2016.2622269}
\showDOI{\tempurl}


\bibitem[Lan et~al\mbox{.}(2021)]%
        {lan21kineticharts}
\bibfield{author}{\bibinfo{person}{Xingyu Lan}, \bibinfo{person}{Yang Shi},
  \bibinfo{person}{Yanqiu Wu}, \bibinfo{person}{Xiaohan Jiao}, {and}
  \bibinfo{person}{Nan Cao}.} \bibinfo{year}{2021}\natexlab{}.
\newblock \showarticletitle{Kineticharts: Augmenting Affective Expressiveness
  of Charts in Data Stories with Animation Design}.
\newblock \bibinfo{journal}{\emph{IEEE Transactions on Visualization and
  Computer Graphics}}  \bibinfo{volume}{28} (\bibinfo{year}{2021}),
  \bibinfo{pages}{933--943}.
\newblock
\urldef\tempurl%
\url{https://doi.org/10.1109/TVCG.2021.3114775}
\showDOI{\tempurl}


\bibitem[Lee et~al\mbox{.}(2007)]%
        {Lee07EmotiveCaptioning}
\bibfield{author}{\bibinfo{person}{Daniel~G. Lee}, \bibinfo{person}{Deborah~I.
  Fels}, {and} \bibinfo{person}{John~Patrick Udo}.}
  \bibinfo{year}{2007}\natexlab{}.
\newblock \showarticletitle{Emotive Captioning}.
\newblock \bibinfo{journal}{\emph{Computers in Entertainment}}
  \bibinfo{volume}{5}, \bibinfo{number}{2}, Article \bibinfo{articleno}{11}
  (\bibinfo{year}{2007}), \bibinfo{numpages}{15}~pages.
\newblock
\urldef\tempurl%
\url{https://doi.org/10.1145/1279540.1279551}
\showDOI{\tempurl}


\bibitem[Lee et~al\mbox{.}(2002)]%
        {lee2002engine}
\bibfield{author}{\bibinfo{person}{Johnny~C. Lee}, \bibinfo{person}{Jodi
  Forlizzi}, {and} \bibinfo{person}{Scott~E. Hudson}.}
  \bibinfo{year}{2002}\natexlab{}.
\newblock \showarticletitle{The Kinetic Typography Engine: An Extensible System
  for Animating Expressive Text}. In \bibinfo{booktitle}{\emph{Proceedings of
  the ACM Annual Symposium on User Interface Software and Technology (UIST)}}.
  \bibinfo{publisher}{ACM}, \bibinfo{address}{New York, NY, USA},
  \bibinfo{pages}{81--90}.
\newblock
\urldef\tempurl%
\url{https://doi.org/10.1145/571985.571997}
\showDOI{\tempurl}


\bibitem[Lim(2022)]%
        {lim2022study}
\bibfield{author}{\bibinfo{person}{Sooyeon Lim}.}
  \bibinfo{year}{2022}\natexlab{}.
\newblock \showarticletitle{A Study on the Interactive Expression of Human
  Emotions in Typography}.
\newblock \bibinfo{journal}{\emph{International Journal of Advanced Culture
  Technology}} \bibinfo{volume}{10}, \bibinfo{number}{1}
  (\bibinfo{year}{2022}), \bibinfo{pages}{122--130}.
\newblock
\urldef\tempurl%
\url{https://doi.org/10.17703/IJACT.2022.10.1.122}
\showDOI{\tempurl}


\bibitem[{lipsum.com}(1996)]%
        {lorem}
\bibfield{author}{\bibinfo{person}{{lipsum.com}}.}
  \bibinfo{year}{1996}\natexlab{}.
\newblock \bibinfo{booktitle}{\emph{Lorem Ipsum}}.
\newblock
\urldef\tempurl%
\url{https://www.lipsum.com/}
\showURL{%
Retrieved Mar 20, 2023 from \tempurl}


\bibitem[Malik et~al\mbox{.}(2009)]%
        {malik2009communicating}
\bibfield{author}{\bibinfo{person}{Sabrina Malik}, \bibinfo{person}{Jonathan
  Aitken}, {and} \bibinfo{person}{Judith~Kelly Waalen}.}
  \bibinfo{year}{2009}\natexlab{}.
\newblock \showarticletitle{Communicating Emotion with Animated Text}.
\newblock \bibinfo{journal}{\emph{Visual Communication}} \bibinfo{volume}{8},
  \bibinfo{number}{4} (\bibinfo{year}{2009}), \bibinfo{pages}{469--479}.
\newblock
\urldef\tempurl%
\url{https://doi.org/10.1177/1470357209343375}
\showDOI{\tempurl}


\bibitem[Mao et~al\mbox{.}(2022)]%
        {mao2022intelligent}
\bibfield{author}{\bibinfo{person}{Wendong Mao}, \bibinfo{person}{Shuai Yang},
  \bibinfo{person}{Huihong Shi}, \bibinfo{person}{Jiaying Liu}, {and}
  \bibinfo{person}{Zhongfeng Wang}.} \bibinfo{year}{2022}\natexlab{}.
\newblock \showarticletitle{Intelligent Typography: Artistic Text Style
  Transfer for Complex Texture and Structure}.
\newblock \bibinfo{journal}{\emph{IEEE Transactions on Multimedia}}
  (\bibinfo{year}{2022}).
\newblock
\urldef\tempurl%
\url{https://doi.org/10.1109/TMM.2022.3209870}
\showDOI{\tempurl}
\newblock
\shownote{Early Access}.


\bibitem[Men et~al\mbox{.}(2019)]%
        {men2019dyntypo}
\bibfield{author}{\bibinfo{person}{Yifang Men}, \bibinfo{person}{Zhouhui Lian},
  \bibinfo{person}{Yingmin Tang}, {and} \bibinfo{person}{Jianguo Xiao}.}
  \bibinfo{year}{2019}\natexlab{}.
\newblock \showarticletitle{DynTypo: Example-based Dynamic Text Effects
  Transfer}. In \bibinfo{booktitle}{\emph{Proceedings of the IEEE/CVF
  Conference on Computer Vision and Pattern Recognition}}.
  \bibinfo{publisher}{IEEE}, \bibinfo{address}{Piscataway, NJ, USA},
  \bibinfo{pages}{5870--5879}.
\newblock
\urldef\tempurl%
\url{https://doi.org/10.1109/CVPR.2019.00602}
\showDOI{\tempurl}


\bibitem[Minakuchi and Kidawara(2008)]%
        {minakuchi2008kinetic}
\bibfield{author}{\bibinfo{person}{Mitsuru Minakuchi} {and}
  \bibinfo{person}{Yutaka Kidawara}.} \bibinfo{year}{2008}\natexlab{}.
\newblock \showarticletitle{Kinetic Typography for Ambient Displays}. In
  \bibinfo{booktitle}{\emph{Proceedings of the International Conference on
  Ubiquitous Information Management and Communication (ICUIMC)}}.
  \bibinfo{publisher}{ACM}, \bibinfo{address}{New York, NY, USA},
  \bibinfo{pages}{54--57}.
\newblock
\urldef\tempurl%
\url{https://doi.org/10.1145/1352793.1352805}
\showDOI{\tempurl}


\bibitem[Minakuchi and Tanaka(2005)]%
        {minakuchi2005kinetic}
\bibfield{author}{\bibinfo{person}{Mitsuru Minakuchi} {and}
  \bibinfo{person}{Katsumi Tanaka}.} \bibinfo{year}{2005}\natexlab{}.
\newblock \showarticletitle{Automatic Kinetic Typography Composer}. In
  \bibinfo{booktitle}{\emph{Proceedings of the ACM Conference on Advances in
  Computer Entertainment Technology (ACE)}}. \bibinfo{publisher}{ACM},
  \bibinfo{address}{New York, NY, USA}, \bibinfo{pages}{221--224}.
\newblock
\showISBNx{1595931104}
\urldef\tempurl%
\url{https://doi.org/10.1145/1178477.1178512}
\showDOI{\tempurl}


\bibitem[{Moonlab Studio Co., Ltd}(2023)]%
        {linedog}
\bibfield{author}{\bibinfo{person}{{Moonlab Studio Co., Ltd}}.}
  \bibinfo{year}{2023}\natexlab{}.
\newblock \bibinfo{booktitle}{\emph{Puppy Maltese}}.
\newblock
\urldef\tempurl%
\url{https://weibo.com/u/7776232700?tabtype=home}
\showURL{%
Retrieved Mar 20, 2023 from \tempurl}


\bibitem[Penny(1996)]%
        {penney1996truetype}
\bibfield{author}{\bibinfo{person}{Laurence Penny}.}
  \bibinfo{year}{1996}\natexlab{}.
\newblock \bibinfo{booktitle}{\emph{A History of TrueType}}.
\newblock
\urldef\tempurl%
\url{https://www.truetype-typography.com}
\showURL{%
Retrieved Mar 20, 2023 from \tempurl}


\bibitem[Phan et~al\mbox{.}(2015)]%
        {phan2015flexyfont}
\bibfield{author}{\bibinfo{person}{Huy~Quoc Phan}, \bibinfo{person}{Hongbo Fu},
  {and} \bibinfo{person}{Antoni~B Chan}.} \bibinfo{year}{2015}\natexlab{}.
\newblock \showarticletitle{Flexyfont: Learning Transferring Rules for Flexible
  Typeface Synthesis}.
\newblock \bibinfo{journal}{\emph{Computer Graphics Forum}}
  \bibinfo{volume}{34}, \bibinfo{number}{7} (\bibinfo{year}{2015}),
  \bibinfo{pages}{245--256}.
\newblock
\urldef\tempurl%
\url{https://doi.org/10.1111/cgf.12763}
\showDOI{\tempurl}


\bibitem[Rombach et~al\mbox{.}(2022)]%
        {rombach2022high}
\bibfield{author}{\bibinfo{person}{Robin Rombach}, \bibinfo{person}{Andreas
  Blattmann}, \bibinfo{person}{Dominik Lorenz}, \bibinfo{person}{Patrick
  Esser}, {and} \bibinfo{person}{Bj{\"o}rn Ommer}.}
  \bibinfo{year}{2022}\natexlab{}.
\newblock \showarticletitle{High-Resolution Image Synthesis with Latent
  Diffusion Models}. In \bibinfo{booktitle}{\emph{Proceedings of the IEEE/CVF
  Conference on Computer Vision and Pattern Recognition (CVPR)}}.
  \bibinfo{publisher}{IEEE}, \bibinfo{address}{Piscataway, NJ, USA},
  \bibinfo{pages}{10684--10695}.
\newblock
\urldef\tempurl%
\url{https://doi.org/10.1109/CVPR52688.2022.01042}
\showDOI{\tempurl}


\bibitem[Shu et~al\mbox{.}(2021)]%
        {Shu21Gif}
\bibfield{author}{\bibinfo{person}{Xinhuan Shu}, \bibinfo{person}{Aoyu Wu},
  \bibinfo{person}{Junxiu Tang}, \bibinfo{person}{Benjamin Bach},
  \bibinfo{person}{Yingcai Wu}, {and} \bibinfo{person}{Huamin Qu}.}
  \bibinfo{year}{2021}\natexlab{}.
\newblock \showarticletitle{What Makes a {Data-GIF} Understandable?}
\newblock \bibinfo{journal}{\emph{IEEE Trans. Vis. Comput. Graph.}}
  \bibinfo{volume}{27}, \bibinfo{number}{02} (\bibinfo{year}{2021}),
  \bibinfo{pages}{1492--1502}.
\newblock
\urldef\tempurl%
\url{https://doi.org/10.1109/TVCG.2020.3030396}
\showDOI{\tempurl}


\bibitem[Siarohin et~al\mbox{.}(2019a)]%
        {siarohin2019animating}
\bibfield{author}{\bibinfo{person}{Aliaksandr Siarohin},
  \bibinfo{person}{St{\'e}phane Lathuili{\`e}re}, \bibinfo{person}{Sergey
  Tulyakov}, \bibinfo{person}{Elisa Ricci}, {and} \bibinfo{person}{Nicu Sebe}.}
  \bibinfo{year}{2019}\natexlab{a}.
\newblock \showarticletitle{Animating Arbitrary Objects via Deep Motion
  Transfer}. In \bibinfo{booktitle}{\emph{Proceedings of the IEEE/CVF
  Conference on Computer Vision and Pattern Recognition (CVPR)}}.
  \bibinfo{publisher}{IEEE}, \bibinfo{address}{Piscataway, NJ, USA},
  \bibinfo{pages}{2377--2386}.
\newblock
\urldef\tempurl%
\url{https://doi.org/10.1109/CVPR.2019.00248}
\showDOI{\tempurl}


\bibitem[Siarohin et~al\mbox{.}(2019b)]%
        {siarohin2019first}
\bibfield{author}{\bibinfo{person}{Aliaksandr Siarohin},
  \bibinfo{person}{St{\'e}phane Lathuili{\`e}re}, \bibinfo{person}{Sergey
  Tulyakov}, \bibinfo{person}{Elisa Ricci}, {and} \bibinfo{person}{Nicu Sebe}.}
  \bibinfo{year}{2019}\natexlab{b}.
\newblock \showarticletitle{First Order Motion Model for Image Animation}. In
  \bibinfo{booktitle}{\emph{Proceedings of the International Conference on
  Neural Information Processing Systems (NeurIPs)}}. \bibinfo{publisher}{Curran
  Associates Inc.}, \bibinfo{address}{Red Hook, NY, USA}, Article
  \bibinfo{articleno}{641}, \bibinfo{numpages}{11}~pages.
\newblock
\urldef\tempurl%
\url{https://doi.org/10.5555/3454287.3454928}
\showDOI{\tempurl}


\bibitem[Siarohin et~al\mbox{.}(2021)]%
        {siarohin2021motion}
\bibfield{author}{\bibinfo{person}{Aliaksandr Siarohin},
  \bibinfo{person}{Oliver~J Woodford}, \bibinfo{person}{Jian Ren},
  \bibinfo{person}{Menglei Chai}, {and} \bibinfo{person}{Sergey Tulyakov}.}
  \bibinfo{year}{2021}\natexlab{}.
\newblock \showarticletitle{Motion Representations for Articulated Animation}.
  In \bibinfo{booktitle}{\emph{Proceedings of the IEEE/CVF Conference on
  Computer Vision and Pattern Recognition (CVPR)}}. \bibinfo{publisher}{IEEE},
  \bibinfo{address}{Piscataway, NJ, USA}, \bibinfo{pages}{13653--13662}.
\newblock
\urldef\tempurl%
\url{https://doi.org/10.1109/CVPR46437.2021.01344}
\showDOI{\tempurl}


\bibitem[Smith et~al\mbox{.}(2023)]%
        {smith2023tog}
\bibfield{author}{\bibinfo{person}{Harrison~Jesse Smith},
  \bibinfo{person}{Qingyuan Zheng}, \bibinfo{person}{Yifei Li},
  \bibinfo{person}{Somya Jain}, {and} \bibinfo{person}{Jessica~K. Hodgins}.}
  \bibinfo{year}{2023}\natexlab{}.
\newblock \showarticletitle{A Method for Animating Children’s Drawings of the
  Human Figure}.
\newblock \bibinfo{journal}{\emph{ACM Transactions on Graphics}}
  \bibinfo{volume}{42}, \bibinfo{number}{3}, Article \bibinfo{articleno}{32}
  (\bibinfo{year}{2023}), \bibinfo{numpages}{15}~pages.
\newblock
\showISSN{0730-0301}
\urldef\tempurl%
\url{https://doi.org/10.1145/3592788}
\showDOI{\tempurl}


\bibitem[Su et~al\mbox{.}(2018)]%
        {su2018live}
\bibfield{author}{\bibinfo{person}{Qingkun Su}, \bibinfo{person}{Xue Bai},
  \bibinfo{person}{Hongbo Fu}, \bibinfo{person}{Chiew-Lan Tai}, {and}
  \bibinfo{person}{Jue Wang}.} \bibinfo{year}{2018}\natexlab{}.
\newblock \showarticletitle{Live Sketch: Video-Driven Dynamic Deformation of
  Static Drawings}. In \bibinfo{booktitle}{\emph{Proceedings of the ACM
  Conference on Human Factors in Computing Systems (CHI)}}.
  \bibinfo{publisher}{Association for Computing Machinery},
  \bibinfo{address}{New York, NY, USA}, Article \bibinfo{articleno}{662},
  \bibinfo{numpages}{12}~pages.
\newblock
\showISBNx{9781450356206}
\urldef\tempurl%
\url{https://doi.org/10.1145/3173574.3174236}
\showDOI{\tempurl}


\bibitem[Tao et~al\mbox{.}(2022)]%
        {tao2022structure}
\bibfield{author}{\bibinfo{person}{Jiale Tao}, \bibinfo{person}{Biao Wang},
  \bibinfo{person}{Borun Xu}, \bibinfo{person}{Tiezheng Ge},
  \bibinfo{person}{Yuning Jiang}, \bibinfo{person}{Wen Li}, {and}
  \bibinfo{person}{Lixin Duan}.} \bibinfo{year}{2022}\natexlab{}.
\newblock \showarticletitle{Structure-Aware Motion Transfer with Deformable
  Anchor Model}. In \bibinfo{booktitle}{\emph{Proceedings of the IEEE/CVF
  Conference on Computer Vision and Pattern Recognition (CVPR)}}.
  \bibinfo{publisher}{IEEE}, \bibinfo{address}{Piscataway, NJ, USA},
  \bibinfo{pages}{3637--3646}.
\newblock
\urldef\tempurl%
\url{https://doi.org/10.1109/CVPR52688.2022.00362}
\showDOI{\tempurl}


\bibitem[Tendulkar et~al\mbox{.}(2019)]%
        {tendulkar2019trick}
\bibfield{author}{\bibinfo{person}{Purva Tendulkar}, \bibinfo{person}{Kalpesh
  Krishna}, \bibinfo{person}{Ramprasaath~R Selvaraju}, {and}
  \bibinfo{person}{Devi Parikh}.} \bibinfo{year}{2019}\natexlab{}.
\newblock \showarticletitle{Trick or TReAT: Thematic Reinforcement for Artistic
  Typography}. In \bibinfo{booktitle}{\emph{Proceedings of the International
  Conferences on Computational Creativity (ICCC)}}. \bibinfo{publisher}{ACC},
  \bibinfo{numpages}{9}~pages.
\newblock
\showeprint{1903.07820}


\bibitem[{The Walt Disney Company}(1940)]%
        {fantasia}
\bibfield{author}{\bibinfo{person}{{The Walt Disney Company}}.}
  \bibinfo{year}{1940}\natexlab{}.
\newblock \bibinfo{booktitle}{\emph{Fantasia}}.
\newblock
\urldef\tempurl%
\url{https://youtu.be/r7gLlIv4ito}
\showURL{%
Retrieved Mar 20, 2023 from \tempurl}
\newblock
\shownote{17:03--18:07}.


\bibitem[Thomas et~al\mbox{.}(1995)]%
        {thomas1995illusion}
\bibfield{author}{\bibinfo{person}{Frank Thomas}, \bibinfo{person}{Ollie
  Johnston}, {and} \bibinfo{person}{Frank Thomas}.}
  \bibinfo{year}{1995}\natexlab{}.
\newblock \bibinfo{booktitle}{\emph{The Illusion of Life: Disney Animation}}.
\newblock \bibinfo{publisher}{Hyperion New York}.
\newblock


\bibitem[Vi{\'{e}}gas et~al\mbox{.}(2009)]%
        {viegas09participatory}
\bibfield{author}{\bibinfo{person}{Fernanda~B. Vi{\'{e}}gas},
  \bibinfo{person}{Martin Wattenberg}, {and} \bibinfo{person}{Jonathan
  Feinberg}.} \bibinfo{year}{2009}\natexlab{}.
\newblock \showarticletitle{Participatory Visualization with Wordle}.
\newblock \bibinfo{journal}{\emph{IEEE Transactions on Visualization and
  Computer Graphics}} \bibinfo{volume}{15}, \bibinfo{number}{6}
  (\bibinfo{year}{2009}), \bibinfo{pages}{1137--1144}.
\newblock
\urldef\tempurl%
\url{https://doi.org/10.1109/TVCG.2009.171}
\showDOI{\tempurl}


\bibitem[Vy et~al\mbox{.}(2008)]%
        {vy2008enact}
\bibfield{author}{\bibinfo{person}{Quoc~V Vy}, \bibinfo{person}{Jorge~A Mori},
  \bibinfo{person}{David~W Fourney}, {and} \bibinfo{person}{Deborah~I Fels}.}
  \bibinfo{year}{2008}\natexlab{}.
\newblock \showarticletitle{EnACT: A Software Tool for Creating Animated Text
  captions}. In \bibinfo{booktitle}{\emph{Proceedings of the Internation
  Conference on Computers Helping People with Special Needs (ICCHP)}}.
  \bibinfo{publisher}{Springer}, \bibinfo{address}{Berlin, Heidelberg},
  \bibinfo{pages}{609--616}.
\newblock
\urldef\tempurl%
\url{https://doi.org/10.1007/978-3-540-70540-6_87}
\showDOI{\tempurl}


\bibitem[Wang et~al\mbox{.}(2004)]%
        {Wang2004communicate}
\bibfield{author}{\bibinfo{person}{Hua Wang}, \bibinfo{person}{Helmut
  Prendinger}, {and} \bibinfo{person}{Takeo Igarashi}.}
  \bibinfo{year}{2004}\natexlab{}.
\newblock \showarticletitle{Communicating Emotions in Online Chat Using
  Physiological Sensors and Animated Text}. In
  \bibinfo{booktitle}{\emph{Extended Abstracts on Human Factors in Computing
  Systems (CHIEA)}}. \bibinfo{publisher}{ACM}, \bibinfo{address}{New York, NY,
  USA}, \bibinfo{pages}{1171--1174}.
\newblock
\urldef\tempurl%
\url{https://doi.org/10.1145/985921.986016}
\showDOI{\tempurl}


\bibitem[Wang et~al\mbox{.}(2019)]%
        {wang2019typography}
\bibfield{author}{\bibinfo{person}{Wenjing Wang}, \bibinfo{person}{Jiaying
  Liu}, \bibinfo{person}{Shuai Yang}, {and} \bibinfo{person}{Zongming Guo}.}
  \bibinfo{year}{2019}\natexlab{}.
\newblock \showarticletitle{Typography with Decor: Intelligent Text Style
  Transfer}. In \bibinfo{booktitle}{\emph{Proceedings of the IEEE/CVF
  Conference on Computer Vision and Pattern Recognition (CVPR)}}.
  \bibinfo{publisher}{IEEE}, \bibinfo{address}{Piscataway, NJ, USA},
  \bibinfo{pages}{5889--5897}.
\newblock
\urldef\tempurl%
\url{https://doi.org/10.1109/CVPR.2019.00604}
\showDOI{\tempurl}


\bibitem[Wang et~al\mbox{.}(2021)]%
        {wang2021infomotion}
\bibfield{author}{\bibinfo{person}{Yun Wang}, \bibinfo{person}{Yi Gao},
  \bibinfo{person}{Ray Huang}, \bibinfo{person}{Weiwei Cui},
  \bibinfo{person}{Haidong Zhang}, {and} \bibinfo{person}{Dongmei Zhang}.}
  \bibinfo{year}{2021}\natexlab{}.
\newblock \showarticletitle{Animated Presentation of Static Infographics with
  {InfoMotion}}.
\newblock \bibinfo{journal}{\emph{Computer Graphics Forum}}
  \bibinfo{volume}{40}, \bibinfo{number}{3} (\bibinfo{year}{2021}),
  \bibinfo{pages}{507--518}.
\newblock
\urldef\tempurl%
\url{https://doi.org/10.1111/cgf.14325}
\showDOI{\tempurl}


\bibitem[Willett et~al\mbox{.}(2018)]%
        {willett2018mixed}
\bibfield{author}{\bibinfo{person}{Nora~S. Willett},
  \bibinfo{person}{Rubaiat~Habib Kazi}, \bibinfo{person}{Michael Chen},
  \bibinfo{person}{George Fitzmaurice}, \bibinfo{person}{Adam Finkelstein},
  {and} \bibinfo{person}{Tovi Grossman}.} \bibinfo{year}{2018}\natexlab{}.
\newblock \showarticletitle{A Mixed-Initiative Interface for Animating Static
  Pictures}. In \bibinfo{booktitle}{\emph{Proceedings of the ACM Annual
  Symposium on User Interface Software and Technology (UIST)}}.
  \bibinfo{publisher}{ACM}, \bibinfo{address}{New York, NY, USA},
  \bibinfo{pages}{649--661}.
\newblock
\urldef\tempurl%
\url{https://doi.org/10.1145/3242587.3242612}
\showDOI{\tempurl}


\bibitem[Willett et~al\mbox{.}(2020)]%
        {willett2020pose2pose}
\bibfield{author}{\bibinfo{person}{Nora~S Willett},
  \bibinfo{person}{Hijung~Valentina Shin}, \bibinfo{person}{Zeyu Jin},
  \bibinfo{person}{Wilmot Li}, {and} \bibinfo{person}{Adam Finkelstein}.}
  \bibinfo{year}{2020}\natexlab{}.
\newblock \showarticletitle{Pose2Pose: Pose Selection and Transfer for 2D
  Character Animation}. In \bibinfo{booktitle}{\emph{Proceedings of the 25th
  International Conference on Intelligent User Interfaces (IUI)}}.
  \bibinfo{publisher}{ACM}, \bibinfo{address}{New York, NY, USA},
  \bibinfo{pages}{88--99}.
\newblock
\urldef\tempurl%
\url{https://doi.org/10.1145/3377325.3377505}
\showDOI{\tempurl}


\bibitem[Xie et~al\mbox{.}(2023)]%
        {xie2023emordle}
\bibfield{author}{\bibinfo{person}{Liwenhan Xie}, \bibinfo{person}{Xinhuan
  Shu}, \bibinfo{person}{Jeon~Cheol Su}, \bibinfo{person}{Yun Wang},
  \bibinfo{person}{Siming Chen}, {and} \bibinfo{person}{Huamin Qu}.}
  \bibinfo{year}{2023}\natexlab{}.
\newblock \showarticletitle{Creating Emordle: Animating Word Cloud for Emotion
  Expression}.
\newblock \bibinfo{journal}{\emph{IEEE Transactions on Visualization and
  Computer Graphics}} (\bibinfo{year}{2023}).
\newblock
\urldef\tempurl%
\url{https://doi.org/10.1109/TVCG.2023.3286392}
\showDOI{\tempurl}
\newblock
\shownote{Early Access}.


\bibitem[Xing et~al\mbox{.}(2016)]%
        {xing2016energy}
\bibfield{author}{\bibinfo{person}{Jun Xing}, \bibinfo{person}{Rubaiat~Habib
  Kazi}, \bibinfo{person}{Tovi Grossman}, \bibinfo{person}{Li-Yi Wei},
  \bibinfo{person}{Jos Stam}, {and} \bibinfo{person}{George Fitzmaurice}.}
  \bibinfo{year}{2016}\natexlab{}.
\newblock \showarticletitle{Energy-Brushes: Interactive Tools for Illustrating
  Stylized Elemental Dynamics}. In \bibinfo{booktitle}{\emph{Proceedings of the
  ACM Annual Symposium on User Interface Software and Technology (UIST)}}.
  \bibinfo{publisher}{ACM}, \bibinfo{address}{New York, NY, USA},
  \bibinfo{pages}{755--766}.
\newblock
\urldef\tempurl%
\url{https://doi.org/10.1145/2984511.2984585}
\showDOI{\tempurl}


\bibitem[Xu et~al\mbox{.}(2022)]%
        {xu2022motion}
\bibfield{author}{\bibinfo{person}{Borun Xu}, \bibinfo{person}{Biao Wang},
  \bibinfo{person}{Jinhong Deng}, \bibinfo{person}{Jiale Tao},
  \bibinfo{person}{Tiezheng Ge}, \bibinfo{person}{Yuning Jiang},
  \bibinfo{person}{Wen Li}, {and} \bibinfo{person}{Lixin Duan}.}
  \bibinfo{year}{2022}\natexlab{}.
\newblock \showarticletitle{Motion and Appearance Adaptation for Cross-Domain
  Motion Transfer}. In \bibinfo{booktitle}{\emph{Proceedings of the European
  Conference on Computer Vision (ECCV), Part XVI}}.
  \bibinfo{publisher}{Springer}, \bibinfo{address}{Cham, Germany},
  \bibinfo{pages}{529--545}.
\newblock
\urldef\tempurl%
\url{https://doi.org/10.1007/978-3-031-19787-1_40}
\showDOI{\tempurl}


\bibitem[Xu and Kaplan(2007)]%
        {xu2007calligraphic}
\bibfield{author}{\bibinfo{person}{Jie Xu} {and} \bibinfo{person}{Craig~S
  Kaplan}.} \bibinfo{year}{2007}\natexlab{}.
\newblock \showarticletitle{Calligraphic Packing}. In
  \bibinfo{booktitle}{\emph{Proceedings of Graphics Interface (GI)}}.
  \bibinfo{publisher}{ACM}, \bibinfo{address}{New York, NY, USA},
  \bibinfo{pages}{43--50}.
\newblock
\urldef\tempurl%
\url{https://doi.org/10.1145/1268517.1268527}
\showDOI{\tempurl}


\bibitem[Xu et~al\mbox{.}(2008)]%
        {xu2008animating}
\bibfield{author}{\bibinfo{person}{Xuemiao Xu}, \bibinfo{person}{Liang Wan},
  \bibinfo{person}{Xiaopei Liu}, \bibinfo{person}{Tien-Tsin Wong},
  \bibinfo{person}{Liansheng Wang}, {and} \bibinfo{person}{Chi-Sing Leung}.}
  \bibinfo{year}{2008}\natexlab{}.
\newblock \showarticletitle{Animating Animal Motion from Still}. In
  \bibinfo{booktitle}{\emph{Proceedings of the Special Interest Group on
  Computer Graphics and Interactive Techniques (SIGGRAPH Asia)}}.
  \bibinfo{publisher}{ACM}, \bibinfo{address}{New York, NY, USA}, Article
  \bibinfo{articleno}{117}, \bibinfo{numpages}{8}~pages.
\newblock
\urldef\tempurl%
\url{https://doi.org/10.1145/1457515.1409070}
\showDOI{\tempurl}


\bibitem[Yang et~al\mbox{.}(2021)]%
        {yang2021shape}
\bibfield{author}{\bibinfo{person}{Shuai Yang}, \bibinfo{person}{Zhangyang
  Wang}, {and} \bibinfo{person}{Jiaying Liu}.} \bibinfo{year}{2021}\natexlab{}.
\newblock \showarticletitle{Shape-Matching GAN++: Scale Controllable Dynamic
  Artistic Text Style Transfer}.
\newblock \bibinfo{journal}{\emph{IEEE Transactions on Pattern Analysis and
  Machine Intelligence}} \bibinfo{volume}{44}, \bibinfo{number}{7}
  (\bibinfo{year}{2021}), \bibinfo{pages}{3807--3820}.
\newblock
\urldef\tempurl%
\url{https://doi.org/10.1109/TPAMI.2021.3055211}
\showDOI{\tempurl}


\bibitem[Yeo(2008)]%
        {yeo2008kim}
\bibfield{author}{\bibinfo{person}{Zhiquan Yeo}.}
  \bibinfo{year}{2008}\natexlab{}.
\newblock \showarticletitle{Emotional Instant Messaging with KIM}. In
  \bibinfo{booktitle}{\emph{Extended Abstracts on Human Factors in Computing
  Systems (CHIEA)}}. \bibinfo{publisher}{ACM}, \bibinfo{address}{New York, NY,
  USA}, \bibinfo{pages}{3729--3734}.
\newblock
\urldef\tempurl%
\url{https://doi.org/10.1145/1358628.1358921}
\showDOI{\tempurl}


\bibitem[Zhang et~al\mbox{.}(2017)]%
        {zhang2017synthesizing}
\bibfield{author}{\bibinfo{person}{Junsong Zhang}, \bibinfo{person}{Yu Wang},
  \bibinfo{person}{Weiyi Xiao}, {and} \bibinfo{person}{Zhenshan Luo}.}
  \bibinfo{year}{2017}\natexlab{}.
\newblock \showarticletitle{Synthesizing Ornamental Typefaces}.
\newblock \bibinfo{journal}{\emph{Computer Graphics Forum}}
  \bibinfo{volume}{36}, \bibinfo{number}{1} (\bibinfo{year}{2017}),
  \bibinfo{pages}{64--75}.
\newblock
\urldef\tempurl%
\url{https://doi.org/10.1111/cgf.12785}
\showDOI{\tempurl}


\bibitem[Zhou et~al\mbox{.}(2020)]%
        {zhou2020makeittalk}
\bibfield{author}{\bibinfo{person}{Yang Zhou}, \bibinfo{person}{Xintong Han},
  \bibinfo{person}{Eli Shechtman}, \bibinfo{person}{Jose Echevarria},
  \bibinfo{person}{Evangelos Kalogerakis}, {and} \bibinfo{person}{Dingzeyu
  Li}.} \bibinfo{year}{2020}\natexlab{}.
\newblock \showarticletitle{MakeltTalk: Speaker-Aware Talking-Head Animation}.
\newblock \bibinfo{journal}{\emph{ACM Transactions on Graphics}}
  \bibinfo{volume}{39}, \bibinfo{number}{6}, Article \bibinfo{articleno}{221}
  (\bibinfo{year}{2020}), \bibinfo{numpages}{15}~pages.
\newblock
\urldef\tempurl%
\url{https://doi.org/10.1145/3414685.3417774}
\showDOI{\tempurl}


\bibitem[Zou et~al\mbox{.}(2016)]%
        {zou2016legible}
\bibfield{author}{\bibinfo{person}{Changqing Zou}, \bibinfo{person}{Junjie
  Cao}, \bibinfo{person}{Warunika Ranaweera}, \bibinfo{person}{Ibraheem
  Alhashim}, \bibinfo{person}{Ping Tan}, \bibinfo{person}{Alla Sheffer}, {and}
  \bibinfo{person}{Hao Zhang}.} \bibinfo{year}{2016}\natexlab{}.
\newblock \showarticletitle{Legible Compact Calligrams}.
\newblock \bibinfo{journal}{\emph{ACM Transactions on Graphics}}
  \bibinfo{volume}{35}, \bibinfo{number}{4}, Article \bibinfo{articleno}{122}
  (\bibinfo{year}{2016}), \bibinfo{numpages}{12}~pages.
\newblock
\urldef\tempurl%
\url{https://doi.org/10.1145/2897824.2925887}
\showDOI{\tempurl}


\bibitem[Zucker(2023)]%
        {typemonkey}
\bibfield{author}{\bibinfo{person}{Ebberts~+ Zucker}.}
  \bibinfo{year}{2023}\natexlab{}.
\newblock \bibinfo{booktitle}{\emph{TypeMonkey}}.
\newblock
\urldef\tempurl%
\url{http://aescripts.com/typemonkey/}
\showURL{%
Retrieved Mar 20, 2023 from \tempurl}


\end{thebibliography}
